\documentclass[twocolumn, prb,preprintnumbers,amsmath,amssymb,floatfix]{revtex4}
\usepackage[linktocpage,bookmarksopen,bookmarksnumbered]{hyperref}

\usepackage{graphicx}
\usepackage{dcolumn}

\usepackage{amsmath,graphics,epsfig,color,verbatim,ulem}
\usepackage{braket}
\newcommand{\eps}{\epsilon} 
 \newcommand{\vR}{{\mathbf{R}}}
\renewcommand{\vr}{{\mathbf{r}}} 
\newcommand{\vk}{{\mathbf{k}}}
\newcommand{\vb}{{\mathbf{b}}}

\newcommand{\vp}{{\mathbf{p}}}
\newcommand{\vG}{{\mathbf{G}}}
 \newcommand{\vq}{{\mathbf{q}}}

\usepackage{multirow}

\usepackage{tabularx,ragged2e,booktabs,caption}
\usepackage{cancel}
\newcolumntype{C}[1]{>{\Centering}m{#1}}

\usepackage{multibib}

\begin{document}

\setlength{\pdfpageheight}{\paperheight}
\setlength{\pdfpagewidth}{\paperwidth}


\title{All electron GW with linearized augmented plane waves for metals and semiconductors}
\author{Kristjan Haule}

\address{Center for Materials Theory and Department of Physics and Astronomy, Rutgers University, Piscataway, NJ 08854, United States.}
\author{Subhasish Mandal}
\address{Department of Physics and Astronomy, West Virginia University, Morgantown, WV 26506, United States.}

\begin{abstract}
GW approximation is one of the most popular parameter-free many-body methods that go beyond the limitations of the standard density functional theory (DFT) to determine the excitation spectra for moderately correlated materials and in particular the semiconductors. It is also the first step in developing the diagrammatic Monte Carlo method into an electronic structure tool, which would offer a numerically exact solution to the solid-state problem. While most electronic structure packages offer support for GW calculations for band-insulating materials, the level of support for metallic systems is somewhat limited. This limitation can be partly attributed to the relatively minor differences often observed between GW and DFT results in treating metallic systems, which is not expected to persist to higher orders in perturbation theory.
Describing metals within the GW framework presents a challenge, as it requires accurate resolution of Fermi surface singularities, which, in turn, calls for a dense momentum mesh.
Here we implement the GW algorithm within the all-electron Linear Augmented Plane Wave framework, where we pay special attention to the metallic systems, the convergence with respect to momentum mesh, and proper treatment of the deep laying core states, as needed for the future variational diagrammatic Monte Carlo implementation.
Our improved algorithm for resolving Fermi surface singularities allows us a stable and accurate analytic continuation of imaginary axis data, which is carried out for GW excitation spectra throughout the Brillouin zone in both the metallic and insulating materials and is compared to numerically more stable contour deformation integration technique.   We compute band structures for elemental metallic systems Li, Na, and Mg as well as for various narrow and wide bandgap insulators such as Si, BN, SiC, MgO, LiF, ZnS, and CdS and compare our results with previous GW calculations and available experiments data. Our results are in good agreement with the available literature. Thus our software allows users to compute full bandstructures for metals and insulators using all-electron potential without downfolding to Wannier orbital basis. 
\end{abstract}
\date{\today}
\maketitle


\section{Introduction}
Perturbative expansion around the free electron limit is one of the most common techniques used in the many-body theory. In {\it ab-initio} solid state applications, the expansion is typically carried out in terms of the single-particle Green's function $G$, and the screened Coulomb interaction $W$.
When carried out at the first order approximation, and $W$ is computed by the bubble Feynman diagrams, the method is called the GW approximation~\cite{Hedin}. In  widespread applications of this theory to semiconductors, it was shown that such approximation predicts very accurate band-gaps in semiconductors~\cite{refB4,refB5,refB6,refB7,RMP-GW1,lu_dielectric_2008,rocca_ab_2010} and thus became one of the most popular {\it ab initio} beyond-density functional theory (DFT) approaches in the condensed matter physics and materials science communities.

There were early promising GW studies for weakly interacting metallic systems such as Na~\cite{Hybertsen_Na}, but even 30 years later most electronic structure codes do not offer full support for GW band structure calculation in metallic systems.
%
There are a few notable exceptions, for example
the SPEX code~\cite{Bluegel,referee_4,referee_7,referee_9,referee_11,referee_6},
the ecalj package~\cite{referee_5,referee_8,referee_10,referee_16, referee_18},
and FlapwMBPT code~\cite{Kutepov_Na}.
There are several GW calculations for metals, which used implementations that are not publicly available using pseudopotentials~\cite{referee_13,referee_14,referee_17,referee_12,Metal_study1,Metal_study2} and all electron~\cite{referee_1, referee_2,referee_15,Schilfgaarde1, Schilfgaarde2, Schilfgaarde3, Schilfgaarde4} basis set.
GW calculations for metallic systems remain relatively uncommon when compared to their widespread use in semiconductors. This is due in part to the small differences between GW and DFT, as well as the considerable challenges involved in achieving convergence in GW calculations for metals.
Thus, band structure comparison for metals between angle resolved photoemission spectroscopy (ARPES) experiments and GW calculations are not often seen in the literature, and the convergence of the band structure with momentum mesh is almost never studied. 
Perhaps such slow progress towards GW predictions of band structures in metallic systems is due to the difficulty of resolving the singular excitations around the Fermi surface, which require a large number of momentum points and sophisticated and time consuming analytic contour integration, or stable analytic continuation from the imaginary frequency to the real frequency spectra. Thus it remained  a major challenge to compute accurate band-structure throughout the Brillouin zone for metallic systems using GW approximation, which are converged with respect to the accuracy of the basis set and momentum space mesh. This situation impeded the progress of computational materials design in general.


The accuracy, precision, and scaling of GW calculation, which requires non-local and dynamical self-energy of electron, has considerably improved over the years~\cite{Gap2,SM-GW,rocca_ab_2010,giustino_gw_2010,umari_gw_2010,govoni_large_2015,bruneval_accurate_2008,berger_ab_2010,gao_speeding_2016,liu_cubic_2016,foerster_on3_2011, Bluegel,Blaha_HLO}. On the other hand, better treatment of dynamical self-energy has been achieved in Dynamical Mean Field Theory community~\cite{DMFT_review,Hiroshis,Hiroshi2},
which allows us to reanalyze the predictive power of GW approximation in metallic systems, and perhaps point towards the need of including so-called vertex corrections. 
Recently an alternative point of view to vertex corrections is gaining popularity, namely, Monte Carlo summation of high order Feynman diagrams, which are visited by importance sampling techniques~\cite{Kun,Haule2022,DMC1,DMC2,DMC3,DMC4,DMC5,DMC6,DMC7}. In the quest to develop such a diagrammatic Monte Carlo technique, that can achieve chemical accuracy in solid state applications, very accurate GW implementation with all electron algorithm is needed as the first step. In alternative plane wave implementations, the systematic error due to approximate treatment of core electrons could obscure the improvement brought about by very expensive calculation of the vertex corrections.
The Python implementation of GW developed here~\cite{PyGW_code}, will be used for developing such a systematic diagrammatic Monte Carlo expansion method in the future. As a proof of concept, such high order Feynman expansion method has been recently developed for the simpler but related problem of the electron gas, for which numerically converged results can be obtained in a moderately correlated regime of metallic system~\cite{Kun,Haule2022}, and holds great promise for more widespread applications in solid state systems. 

Here we describe the implementation of GW approximation within the all-electron LAPW framework, paying special attention to metallic systems for which GW calculations are difficult to converge and band structure throughout the Brillouin zone is painful to compute.~\cite{PyGW_code}  We overcame the problem with a more stable implementation of the tetrahedron method, and an improved algorithm for frequency convolution on the Matsubara axis, which allowed us a stable analytic continuation of imaginary axis data by Pade approximation. We crosschecked the Pade analytic continuation by implementing more expensive but more accurate contour deformation integration technique~\cite{contour0a,contour0b,contour1,contour2,contour3}
To produce the band structure plots along the high symmetry direction in momentum space, we implemented two complementary techniques: the interpolation method as described in Refs.~\cite{Pickett_method,Pickett_method0}, as well as wannierization method using maximally localized wannier functions~\cite{PhysRevB.56.12847,RevModPhys.84.1419}.
Finally, we also present a method for numerically  efficient manipulation and storage of Matsubara quantities using  optimized Singular-Value-Decomposition-basis (section \ref{SVD_basis}). 
This package is built upon the Gap2 code~\cite{Gap2_code,Gap2} as a foundation, which also served as the accuracy benchmark at the early stages of development.

This paper is organized as follows. The next section is devoted to the method and presents the setup of perturbation theory in section~\ref{Setup}, followed by the description of the method we use to compute the polarization in Sec.~\ref{Polarization}, and the self-energy in  Sec.~\ref{self-energy}, both
are computed in the eigenbasis of the Coulomb repulsion. In Sec.~\ref{ProductBasis} we discuss the implementation of the product basis, which allows one to write polarization and the Coulomb interaction in two-dimensional matrix form. In Sec.~\ref{SVD_basis} we describe the new algorithm for efficient manipulation of the frequency-dependent quantities $G$ and $W$. Finally, in Sec.~\ref{sef:quasiparticle_dispersion} we present techniques to plot the quasiparticle spectra, from analytic continuation to contour integration technique, and interpolation of band structure using Wannier interpolation as well as minimizing smoothness of bands across the Brillouin zone. We also study the quality of diagonal approximation and compare it to the matrix form of self-energy, and we check the quality of the Pade analytic continuation and compare it to the contour integration technique.

In Sec.~\ref{Results} we compare our implementation of G$_0$W$_0$ and GW$_0$ for insulators with other previously published results, while in Sec.~\ref{Metals} we show band structures of several metals within G$_0$W$_0$ approach. 

\section{Method}

\subsection{Setup of the perturbation theory}
\label{Setup}

Here we concentrate on a diagrammatic point of view of the electronic structure problem, sketching the algorithm in a way in which the extension to higher order diagrams is emphasized, as needed for future Variational Diagrammatic Monte Carlo studies~\cite{Kun}. We mention in passing that our implementation starts from Wien2k implementation~\cite{wien2k} of Kohn-Sham orbitals, and closely follows the algorithm of Gap2 code ~\cite{Gap2,Gap2_code}, and also Ref.~\cite{Bluegel}. Some details can also be found in Ref.~\cite{Blaha_HLO}. Apart from a few bugs found in the Gap2 code, which are detailed here~\cite{PyGW_code}, the output of our PyGW code and Gap2 code is identical for identical input, therefore we managed to reproduce results of Ref. ~\cite{Blaha_HLO}. However, we detail below several improvements of the algorithm, which allows one to treat not only semiconductors but metals as well. 

The building blocks in our setup for the perturbation theory are the Green's functions in the Kohn-Sham basis $G_{\vk,i}=\frac{1}{i\omega+\mu-\varepsilon_{\vk,i}}$, and the Coulomb repulsion is written in its eigenbasis. The former depends on the Kohn-Sham eigenvalues $\varepsilon_{\vk,i}$, while for the latter, we will introduce the so-called product basis~\cite{ProductBasis0}, which is an orthogonal (and overcomplete) basis that faithfully represents products of two Kohn-Sham orbitals, and is here called $\chi_\alpha^\vq(\vr)$. Here $\vr$ stands for the real space vector, and $\vq$ is momentum in the first Brillouin zone. The technical details of how to achieve that within LAPW basis are discussed in section~\ref{ProductBasis}. Once such product basis $\chi_\alpha^\vq(\vr)$  is constructed, we compute the matrix elements between two Kohn-Sham orbitals and this basis functions:
$M_{\alpha,ij} (\vk,\vq)\equiv \braket{\chi^\vq_\alpha| \psi_{\vk,i}\psi^*_{\vk-\vq,j}}$. Similarly, we compute the matrix elements of the Coulomb repulsion on this basis 
$v_{\alpha\beta} (\vq)=\braket{\chi^\vq_\alpha|V_C(\vq)|\chi^\vq_\beta}$, and 
subsequently, we determine the square root of the Coulomb repulsion in its eigenbasis as 
$\sqrt{ v(\vq)}_{\alpha,\beta}= U_{\alpha,l}\sqrt{v_l} \, U^\dagger_{l,\beta}$, where $v_l$ are eigenvalues and $U_{\alpha,l}$ are eigenvectors of the Coulomb repulsion. 

\begin{figure}[bht]
\begin{center}
\includegraphics[width=0.5\linewidth]{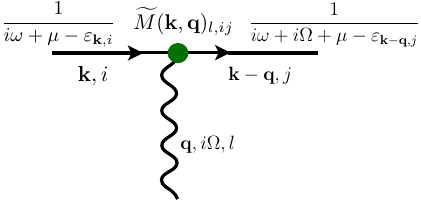}
\caption{ \textbf{Building blocks} of the perturbation theory around DFT starting point. Here $\varepsilon_{\vk,i}$ are energies of the Kohn-Sham orbitals, and $\widetilde{M}(\vk,\vq)$ are matrix elements defined in the text.
}
\label{fig0}
\end{center}
\end{figure}

The interaction between four Kohn-Sham orbitals, in which $\psi_{\vk,i}$, $\psi_{\vk'-\vq,j'}$ are incoming, and $\psi^*_{\vk-\vq,j}$, $\psi_{\vk',i'}^*$ are outgoing electrons, takes the form
\begin{eqnarray}
\psi_{\vk',i'}^*\psi_{\vk'-\vq,j'} v(\vq) \psi_{\vk,i}\psi^*_{\vk-\vq,j}
\end{eqnarray}
and can be evaluated in the product basis by
\begin{eqnarray}
\sum_{\alpha,\beta}\braket{\psi_{\vk',i'}\psi_{\vk'-\vq,j'}^*|\chi^\vq_{\beta}}\braket{\chi^\vq_\beta| {v(\vq)}|\chi^\vq_\alpha} \braket{\chi^\vq_{\alpha}|\psi_{\vk,i}\psi^*_{\vk-\vq,j}}
\end{eqnarray}
which can be expressed with the above-defined matrix elements as
\begin{eqnarray}
\sum_{\alpha,\beta} M_{\beta,i'j'}^*(\vk',\vq)\; v_{\beta\alpha} (\vq)\; M_{\alpha,ij}(\vk,\vq).
\end{eqnarray}
We can now associate a square-root of the Coulomb repulsion with each pair of the Kohn-Sham orbitals and rewrite this product in the above-defined eigenbasis of the Coulomb repulsion as
\begin{eqnarray}
\sum_{l} (M^\dagger(\vk',\vq) U)_{i'j',l} \sqrt{v_l} \sqrt{v_l} (U^{\dagger} M(\vk,\vq))_{l,ij}.
\end{eqnarray}
If we now define the new matrix elements of the form
\begin{eqnarray}
\widetilde{M}(\vk,\vq) \equiv \sqrt{v}\; U^{\dagger} M(\vk,\vq)
\end{eqnarray}
we see that the Coulomb repulsion between the two incoming $\psi_{\vk,i}$, $\psi_{\vk'-\vq,j'}$ and the two outgoing $\psi^*_{\vk-\vq,j}$, $\psi_{\vk',i'}^*$ Kohn-Sham orbitals can in general be written as the product of two matrices
\begin{eqnarray}
\sum_{l} {\widetilde{M}_{i'j',l}^\dagger(\vk',\vq)}  {\widetilde{M}_{l,ij}(\vk,\vq)},
\end{eqnarray}
and hence each three-point vertex can be associated with the matrix element $\widetilde{M}_{l,ij}(\vk,\vq)$, where index $l$ is associated with the bosonic-interaction propagator, and $i$,$j$ with the two Kohn-Sham bands (See Fig.~\ref{fig0}). 
We emphasize that for the perturbative expansion, we only need $\widetilde{M}$ and the Kohn-Sham eigenvalues $\varepsilon_{\vk,i}$ to evaluate the expansion. The matrix elements of the Coulomb repulsion are hence absorbed into the definition of $\widetilde{M}$ and should no longer appear in the calculation.
The advantage of this approach was pointed out in Refs.~\cite{Bluegel}: when the product basis is increased in size so that it becomes more and more precise and complete, there are more and more eigenvalues of the Coulomb repulsion ($v_l$), which are extremely small, and such components can safely be neglected when constructing $\widetilde{M}$. As the linear dimension of the matrix $M$ increases with increasing the energy cutoff for the plane-waves, and the number of radial functions in the spheres, the dimension of $\widetilde{M}$ increases much slower or saturates with increasing the size of the basis. As our calculations only depend on $\widetilde{M}$, this saves a considerable amount of computational time. \\

We want to point out that for the future diagrammatic Monte Carlo calculations, only the matrix $\widetilde{M}_{l,ij}(\vk,\vq)$ will need to be stored, 
apart from Kohn-sham eigenvalues $\varepsilon_{\vk,i}$, to evaluate the Feynman diagram of arbitrary order. However, storing this object in memory will still be a great challenge, as it depends on the dimension of the Coulomb eigenbasis $l$, the square of the number of bands, and also both the fermionic and bosonic momentum. We envision that this matrix $\widetilde{M}$ will need to be stored on a more coarse momentum mesh, and some type of interpolation to a denser mesh of fermionic propagators, which depend on $\varepsilon_{\vk,i}$ and describe the details of the Fermi surface, will need to be implemented.

Finally, let us mention that the single-particle counter-term in this expansion is the Kohn-Sham exchange-correlation potential, which is evaluated in the band-basis by
\begin{eqnarray}
V^{xc}_{ij} = \braket{\psi_{\vk,i}|V^{xc}|\psi_{\vk,j}}
\end{eqnarray}
At the lowest order GW approximation, this potential just needs to be subtracted, and the GW self-energy needs to be added to the Kohn-Sham eigenvalues.
At the higher-order expansion, such a counter-term can be, for example, grouped with the occurrence of exchange sub-diagram in each Feynman diagram, as implemented in Ref.~\cite{Kun}.

\subsection{Polarization}
\label{Polarization}

\begin{figure}[bht]
\begin{center}
\includegraphics[width=0.6\linewidth]{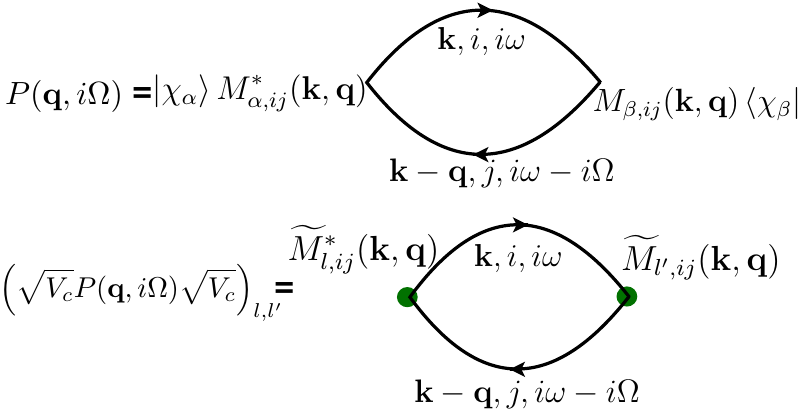}
\caption{ \textbf{Polarization diagram} at the lowest order is the bubble, here expressed in the product basis $\ket{\chi^\vq_\alpha}$. The product of polarization and Coulomb interaction can be expressed in the Coulomb eigenbasis in terms of $\widetilde{M}$ matrix elements only.
}
\label{fig0b}
\end{center}
\end{figure}
The dielectric function in matrix form is $\varepsilon = 1 - \sqrt{V_C} P \sqrt{V_C}$, where $P$ is the polarization.  At the lowest order W$_0$ approximation, the polarization is evaluated as the bubble diagram, which can also be evaluated in the eigenbasis of the Coulomb repulsion, in which it takes the form\\

\begin{widetext}
\begin{eqnarray}
1-\varepsilon_{\vq,i\Omega_n}=(\sqrt{V_C} P \sqrt{V_C})_{l,l'} = N_s\sum_{i,j,\vk}\widetilde{M}^*_{l,ij}(\vk,\vq) \frac{1}{\beta}\sum_{m}\frac{1}{i\omega_m+\mu-\varepsilon_{\vk,i}}\frac{1}{i\omega_m-i\Omega_n+\mu-\varepsilon_{\vk-\vq,j}}\widetilde{M}_{l',ij}(\vk,\vq) 
\nonumber \\
=N_s \sum_{i,j,\vk}\widetilde{M}^*_{l,ij}(\vk,\vq) 
\frac{f(\varepsilon_{\vk-\vq j}-\mu) -f(\varepsilon_{\vk i}-\mu)}{i\Omega_n-\varepsilon_{\vk,i}+\varepsilon_{\vk-\vq,j}}
\widetilde{M}_{l',ij}(\vk,\vq)
\label{Eqeps}
\end{eqnarray}
\end{widetext}
where $f$ is the Fermi function of the form $f(x)=[\exp(x/T)+1]^{-1}$, and indices $i$, $j$ run over Kohn-Sham bands, $N_s$ is 2 or 1 depending on whether the bands contain the spin degeneracy (for example in the presence of the spin-orbit coupling).
It is worth emphasizing that the size of $\varepsilon_{\vq,i\Omega_n}$ matrix is smaller than the size of the product basis, because only the eigenvalues of the Coulomb repulsion ($v_l$), which are finite, contribute to this matrix. Once the matrix $\varepsilon$ is calculated, we invert it in this eigenbasis of the Coulomb repulsion, where the matrix is the smallest.

 In the presence of time reversal symmetry or inversion center, the inner part of the Eq.~\ref{Eqeps} can be rewritten in a more convenient way for computation, such that the band $\vk,i$ is occupied and the $\vk-\vq,j$ is empty, in which case the polarization takes the form
\begin{equation}
P^\vq(i,j,\vk,\Omega_n)=\frac{f(\varepsilon_{\vk i}-\mu) f(-\varepsilon_{\vk-\vq j}+\mu)\; 2(\eps_{\vk i}-\eps_{\vk-\vq j})}{\Omega_n^2 + (\eps_{\vk i}-\eps_{\vk-\vq j})^2}
\label{Eq:even}
\end{equation}
This form emphasizes that the Polarization 
has even symmetry with respect to frequency, and 
is real. However, the matrix elements $\widetilde{M}_{l,ij}(\vk,\vq)$ are in general complex, therefore the polarization is a complex (Hermitian) quantity on the imaginary axis. We use this form for the tetrahedron method, evaluating $\int_{tetra} d^3\vk P^\vq(i,j,\vk,\Omega_n)$, which is implemented similarly as in Gap2 code~\cite{Gap2}, except that we compute all Matsubara frequency points using exactly the same tetrahedron setup, and precompute common parts shared for all Matsubara frequencies, and we group terms which are nearly singular to achieve better cancellation of errors, following ideas from Ref.~\cite{tetra1}, and \cite{tetra0}. In addition, there is a considerable simplification of the tetrahedron method for the case where one of the two bands $i,j$ in the sum is very far from the Fermi level, and therefore only one of the two bands needs to be interpolated, in which case Eq.~\ref{Eq:even} can use the single-particle tetrahedron coefficients, i.e., those that are used to evaluate the densities of states. This reduces the memory requirement in computing the polarization function, as only a limited number of bands around the Fermi level need the sophisticated treatment, while for most of the bands away from the Fermi level, the polarization function Eq.~\ref{Eq:even} can be evaluated on the fly.
The tetrahedron method implemented here, with the innermost loop over Matsubara points, is faster, hence we can afford more Matsubara points. More importantly, the self-energy computed in this way has more uniform frequency dependence, therefore the analytic continuation of the Matsubara self-energy by the standard Pade approximation is now stable, and we can use all computed Matsubara points for Pade analytic continuation, rather than just a couple (for example the two-pole approximation with four Pade coefficients is most common in other implementations \cite{Bluegel,Gap2}).

\subsection{Self-energy}
\label{self-energy}

\begin{figure}[bht]
\begin{center}
\includegraphics[width=0.6\linewidth]{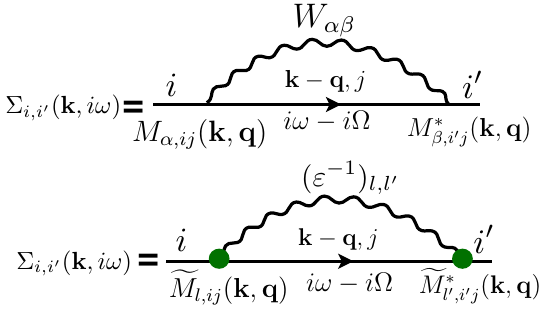}
\caption{ \textbf{The self-energy diagram} at the lowest order GW approximation can also be expressed in terms of matrix elements of $\widetilde{M}$, the dielectric matrix $\varepsilon$, and the single-particle Green's function.
}
\label{fig0c}
\end{center}
\end{figure}
The dynamic correlation self-energy within GW approximation is the convolution of the single-particle Green's function, and the dynamic part of the screened interaction $W-V_C=V_C( \varepsilon^{-1}-1)$, which takes the form
\begin{widetext}
\begin{eqnarray}
\Sigma^c_{ii'}(\vk,i\omega_n)= -\frac{1}{\beta}\sum_{i\Omega_m,\vq,j,\alpha\beta}
\braket{\chi^\vq_\alpha|\psi_{\vk,i}\psi^*_{\vk-\vq,j}}
\braket{\psi_{\vk,i'}\psi^*_{\vk-\vq,j}|\chi^\vq_\beta}\braket{\chi^\vq_\beta|\sqrt{V_C}(\varepsilon^{-1}_{i\Omega_m}-1)\sqrt{V_C}|\chi^\vq_\alpha}G^0_{\vk-\vq,j}(i\omega_n-i\Omega_m)
\nonumber\\
=-\frac{1}{\beta}\sum_{i\Omega_m,\vq,j,l,l'}\widetilde{M}_{l,ij}(\vk,\vq) (\varepsilon^{-1}_{i\Omega_m}-1)_{l,l'} \widetilde{M}^*_{l',i'j}(\vk,\vq) G^0_{\vk-\vq,j}(i\omega_n-i\Omega_m)
\label{Sigma1}
\end{eqnarray}
\end{widetext}
Note that as before, we expressed the self-energy also in terms of the matrix-elements $\widetilde{M}$, written in the eigenbasis of the Coulomb repulsion, which is smaller in dimension than the product basis.
The exchange self-energy is obtained from the above expression by replacing $(\varepsilon^{-1}_{i\Omega_m}-1)_{l,l'}$ with $\delta_{l,l'}$, and it takes the form
\begin{eqnarray}
\Sigma^x_{ii'}(\vk,i\omega_n) =
-\sum_{\vq,j,l}\widetilde{M}_{l,ij}(\vk,\vq) \widetilde{M}^*_{l,i'j}(\vk,\vq) f(\eps_{\vk-\vq,j}-\mu)
\nonumber
\end{eqnarray}

The frequency convolution of the dielectric matrix with the single-particle Green's function can be simplified if we take into account that the polarization is even in frequency $\Omega_m$ (Eq.~\ref{Eq:even}), hence dielectric matrix is also an even function, and therefore
\begin{widetext}
\begin{eqnarray}
\Sigma^c_{ii'}(\vk,i\omega_n)
=-\sum_{\vq,j,l,l'}\widetilde{M}_{l,ij}(\vk,\vq) \widetilde{M}^*_{l',i'j}(\vk,\vq)
\frac{1}{\beta}\sum_{i\Omega_m}
\frac{(\varepsilon^{-1}_{i\Omega_m}-1)_{l,l'} (i\omega_n-\xi_{\vk-\vq,j}-\cancel{i\Omega_m})}{(i\omega_n-\xi_{\vk-\vq,j})^2+\Omega_m^2},
\label{Eq:convol}
\end{eqnarray}
\end{widetext}
i.e., the odd component of the convolution vanishes, and we are left with the sum that falls-off as $1/\Omega_m^4$, because $(\eps^{-1}_{i\Omega_m}-1)$ falls off as $1/\Omega_m^2$. Here $\xi_\vk=\varepsilon_\vk-\mu$. At zero temperature, we can replace the Matsubara sum $\frac{1}{\beta}\sum_{i\Omega_m}$ with the integral $\frac{1}{2\pi}\int_{-\infty}^\infty d\Omega$
hence the inner-convolution in Eq.~\ref{Eq:convol} can be computed by
\begin{eqnarray}
\frac{1}{\pi}\int_0^\infty d\Omega
\frac{(\varepsilon^{-1}_{i\Omega}-1)_{l,l'} (i\omega_n-\xi_{\vk-\vq,j})}{(i\omega_n-\xi_{\vk-\vq,j})^2+\Omega^2}
\label{Eq:convol2}
\end{eqnarray}
To carry out this integral, we spline the quantity $(\varepsilon^{-1}_{i\Omega}-1)(\Omega^2+1)$, which has a nice property that saturates at infinity with vanishing first derivative and also has extremum at zero frequency. We use a vanishing first derivative at infinity and a vanishing second derivative at zero, as the boundary condition for the spline. 
To achieve even better converging integral, we add and subtract a constant such that when $\Omega=\omega_n$ the integrand vanishes. Let us denote $(\varepsilon^{-1}_{i\Omega}-1)_{l,l'} = S_{ll'}(i\Omega)$, then the integral Eq.~\ref{Eq:convol2} can be written as
\begin{eqnarray}
\lim_{L\gg 1}\int_0^L \frac{d\Omega}{\pi}\frac{ S_{ll'}(i\Omega) (\xi_{\vk-\vq,j}-i\omega_n)}{(i\omega_n-\xi_{\vk-\vq,j})^2+\Omega^2}=\\
\lim_{L\gg 1} \frac{\xi_{\vk-\vq,j} -i\omega_n}{\pi} \int_0^L d\Omega\frac{\left(S_{ll'}(i\Omega)-S_{ll'}(i\omega_n)\right) }{(i\omega_n-\xi_{\vk-\vq,j})^2+\Omega^2}
\label{Eq:convol3}
\\+\frac{S_{ll'}(i\omega_n)}{\pi}\arctan{\left(\frac{L}{\xi_{\vk-\vq,j}-i\omega_n}\right)}
\nonumber 
\end{eqnarray}
Using the spline for $S_{ll'}(1+\Omega^2)$, we can afford 10-times or 20-times more frequency points $\Omega$ that the dielectric matrix is calculated on. For both meshes, to compute the dielectric matrix and performing the integral in Eq.~\ref{Eq:convol3}, we use a tangent mesh. This mesh is well-suited for representing Lorentzian function, and is defined by the equation $\Omega = w \tan[x(\pi-2\delta)-\pi/2+\delta]$, where $\delta$ and $w$ are parameters optimized for each represented function, and $x$ is a uniformly spaced mesh in the interval $[-1,1]$.
Here we want to point out that replacing $\arctan{\left(\frac{L}{\xi_{\vk-\vq,j}-i\omega_n}\right)}$ in Eq.~\ref{Eq:convol3} with $\pm\pi/2$ is not precise enough when quantities are known on a finite mesh with cutoff $L$.  
This is because $i\omega$ can also assume large values, resulting in a ratio within the $\arctan$ function that may not necessarily be very large. While it may be tempting to assume that for sufficiently large values of $L$, the values of $S$ would saturate, allowing for the extension of the quadrature to infinity (a practice employed in, for instance, the Gap2 code), our investigations have revealed that results exhibit greater numerical stability when extrapolation is avoided. Instead, employing Matsubara points with a cutoff value consistent with that used in calculating $\varepsilon_{i\Omega}$ yields superior numerical stability.

The correlation self-energy Eq.~\ref{Eq:convol} is
either computed on the Matsubara axis, or directly on the real axis using the contour deformation technique (See section \ref{Sec:contourDef} for details).
When the self-energy is computed on the imaginary axis, it
requires analytic continuation to the real frequency in order to plot band-structure at finite frequency. We managed to implement the tetrahedron method in a stable way so that all Matsubara frequencies $i\Omega_m$ are computed in exactly the same way up to machine precision, therefore we find that standard Pade approximation~\cite{Pade1} is very stable and can be used to plot self-energy on the real axis at frequencies of interest.

\subsection{Product basis within LAPW}
\label{ProductBasis}

The construction of the product basis $\ket{\chi_\alpha}$ has been detailed in prior works, for instance, in~\cite{ProductBasis0,Bluegel,Gap2}. Therefore, here we will provide only a concise summary.
As is customary in the LAPW basis, the space is divided into the muffin-tin (MT) part around each nucleus and the interstitial space in between. Each part of the space has its specific basis functions: plane waves in the interstitial region and radial functions in the MT space. In our implementation, plane waves are utilized exclusively in the interstitial space, while radial functions are employed solely in the MT space. This approach not only facilitates the elimination of linear dependence within the basis but also allows for the use of a more compact product basis.
We note that in our approach the product basis functions $\ket{\chi_\alpha}$ are orthonormal in the MT part, and are also made orthonormal in the interstitial part, which differs from many other implementations, for example Ref.~\cite{ProductBasis0,Bluegel}. We also note that the two parts of the space are treated with its own basis, and therefore functions $\ket{\chi_\alpha}$ are not continuous across the MT-sphere boundary, similarly to most prior implementations~\cite{ProductBasis0,Bluegel,Gap2}. 

In the MT part, the Kohn-Sham wave functions are expanded in terms of the solutions of the radial Schrodinger's equation (at certain energy close to the center of the band) $u_l$, its energy derivative $\dot{u}_l$, and several local orbitals $u^{lo}_l$. Here $l$ is the orbital momentum quantum number. Let's denote all these functions with an index $\kappa$, i.e., $u_l^\kappa$.
The product of the two Kohn-Sham functions spans the Hilbert space which contains all products of such functions $u_l^\kappa u_{l'}^{\kappa'}$. However, we can also order these products in terms of the orbital quantum number $L$ for the products, corresponding to the two-particle orbital momentum $L$. Further, we know that the triangular identity must be satisfied, so that for a given
two-particle momentum $L$ only those single-particle momenta $l$, $l'$ that satisfy $|l-l'|\le L \le l+l'$ can contribute. We can thus construct a limited, yet significant number of products for each $L$, which we denote $\overline{v}_{n,L}$, where $n$ runs over all possible products 
$u_l^\kappa u_{l'}^{\kappa'}$, that satisfy triangular inequality. We then compute overlap between these functions $O_{n,n'}=\braket{\overline{v}_{n,L}|\overline{v}_{n',L}}$ and diagonalize it $O = U \lambda U^\dagger$. Note that here each $L$ is treated independently, and in practice, we can neglect $L$ which are larger than some cutoff (when only $p$ orbitals are occupied, $L=6$ is very accurate, and $L=10$ is converged within a fraction of a percent, hence $2(l+2) \le L \le 2(l+4)$ is good, where $l$ is maximum momentum for occupied single-particle orbital).

The eigenvectors with the eigenvalues larger than some cutoff (for example $10^{-4}$) are assumed to be linearly independent, and are used to construct final product basis functions, i.e.,
\begin{eqnarray}
\ket{v_{\alpha,L}} = \sum_n \ket{\overline{v}_{n,L}} U_{n,\alpha}\frac{1}{\sqrt{\lambda_\alpha}},
\label{BasisDiagonal2}
\end{eqnarray}
where $U$ is defined above as the eigenvector of the overlap ($O = U \lambda U^\dagger$).
Finally, the three dimensional basis functions on the lattice at momentum $\vq$ are constructed with the help of the spherical harmonics:
\begin{eqnarray}
  \braket{\vr\vq|\chi_{\alpha,LM}}_{MT} = v_{\alpha,L}(r) Y_{LM}(\widehat{\vr})
\nonumber
\end{eqnarray}
where $ MT$ means the muffin-tin part of the space. In the interstitial space, we use plane waves of reciprocal vectors $\vG$, i.e.,
\begin{eqnarray}
\braket{\vr\vq|\overline{\chi}_\vG}_{I} = \frac{1}{\sqrt{V}}e^{i(\vq+\vG)\vr}
\nonumber
\end{eqnarray}
where $V$ is the volume of the unit cell. Notice that the Bloch's phase $e^{i\vq\vr}$ is used in the interstitial, but not in the muffin-tin spheres. 

As it is convenient to work with the orthonormal basis, we diagonalize the interstitial basis as well. Just as above we compute the overlap
\begin{eqnarray}
O_{\vG',\vG} = \frac{1}{V_{cell}}\int_{I} e^{i(\vG-\vG')\vr}d^3r \\
= \delta_{GG'}-\sum_{a}\int_{MT_a} e^{i(\vG-\vG')\vr}d^3r
\end{eqnarray}
where $I$ denotes integral over the interstitial space, and $MT_a$ the muffin-tin space of any atom $a$ in the unit cell.
We then diagonalize the overlap $O = U \lambda U^\dagger$, and than construct the orthogonalized plane wave basis as
\begin{eqnarray}
\ket{\chi_\vG}_{I} = \sum_{\vG',\alpha} \ket{\overline{\chi}_{\vG'}}_{I} U_{\vG',\alpha}\frac{1}{\sqrt{\lambda_\alpha}} U^\dagger_{\alpha,\vG}.
\end{eqnarray}
Note that here we added $U^\dagger$ on the right-hand side, as opposed to Eq.~\ref{BasisDiagonal2}, because there is no small eigenvalue in the overlap between plane-waves, and we do not reduce the basis by dropping $U^\dagger$. However, including $U^\dagger$ has a useful effect, namely, the resulting orthogonalized plane waves are gauge invariant, in the sense that they are independent of the arbitrary phase (unitary transformation) of eigenvectors, when diagonalizing complex overlap with many degenerate eigenvalues.

Finally, we want to emphasize that the resulting piece-wise basis, constructed by
\begin{eqnarray}
\ket{\chi} = \left\{
\begin{tabular}{ll}
$\ket{\chi_{\alpha,LM}}$ & $\vr \in$ MT\\
$\ket{\chi_\vG}_{I}$ & $\vr \in$ I
\end{tabular}
\right.
\end{eqnarray}
 is orthonormal, because both parts are orthonormal, and are valid only in their respective parts of the 3D space.
This basis (denoted by $\ket{\chi_\alpha}$) was used in the previous chapter to construct the matrix for the Coulomb repulsion and the dielectric function.

\subsection{SVD frequency basis}
\label{SVD_basis}

We also implemented GW using the minimal frequency basis, obtained by singular-value decomposition of the analytic continuation kernel, the invention of Ref.~\cite{Hiroshis}.
Below we will describe the algorithm in which the frequency dependence of the dielectric matrix can be handled within the minimal basis for bosonic quantities like $W(i\Omega)$. The algorithm was successfully used in the context of Dynamical Mean Field Theory impurity solvers, and in diagrammatic Monte Carlo calculations, but to our knowledge not yet in the context of the GW method. The power of the method is that a very complex imaginary axis function can be represented in terms of a relatively small number of basis functions, and we will show below how to use it to store $W_\vq(i\Omega)$ and speed up the bottleneck of the current GW implementation. However, our current tests show that for materials tested in this report, namely, wide band metals and semiconductors, $W(i\Omega)$ is surprisingly featureless function, and a spline with around 32-64 points on an imaginary axis can describe it with precision around $10^{-10}$. On the other hand, the SVD basis also requires around 30 functions for the same $10^{-10}$ precision, hence we did not manage to achieve considerable speedup with the SVD basis. We note, however, that an SVD basis with 30 functions should be able to describe functions with more complex behavior, in which splines might not perform equally well. The tests on narrow-band metals would probably be more interesting tests of this approach.

The slowest part in our implementation is the computation of the dielectric matrix $\varepsilon$, and in particular its rotation from the band-basis to the product basis. If we denote $P^\vq(i,j,\vk,\Omega)$  in Eq.~\ref{Eq:even} as $p_{ij}(\vk,\vq,i\Omega)$, we can rewrite Eq.~\ref{Eqeps} by
\begin{eqnarray}
(1-\varepsilon)_{ll'} =\sum_{i,j,\vk}\widetilde{M}^*_{l,ij}(\vk,\vq)  p_{ij}(\vk,\vq,i\Omega)\widetilde{M}_{l',ij}(\vk,\vq)
\label{slowest}
\end{eqnarray}
Here $i,j$ are Kohn-Sham band indices, and $l$,$l'$ are Coulomb eigenbasis indices. We note that the dimension of the Coulomb eigenbasis $l$ is substantially smaller than the square of the number of bands, i.e., $i\otimes j$. As this matrix-matrix multiplication takes most of the computational time and needs to be performed for many Matsubara frequencies, it is desirable to find a more compact representation for $p_{ij}(\vk,\vq,i\Omega)$, so that Eq.~\ref{slowest} would need to be performed only a few times. 
The basic idea is to rewrite polarization in the band basis $p_{ij}(\vk,\vq,i\Omega)$ in terms of a small number of svd-basis functions, similarly as in Ref.~\cite{Hiroshis}. The analytic continuation from Matsubara to real frequency is
\begin{eqnarray}
  G(i\Omega) = \int \frac{dx A(x)}{i\Omega-x}
  \label{Esvd1}
\end{eqnarray}
where A(x) is the spectral representation of the correlation function on the real axis. The same equation can be written in discretized form as $G_n = \sum_i K_{n,i}A_i $, where the kernel takes the form: 
\begin{eqnarray}
K_{n,i}\equiv K(\Omega_n, x_i) = \frac{\Delta x_i \sqrt{\Delta \Omega_n}}{i\Omega_n - x_i}
\label{Ker:svd}
\end{eqnarray}
and $\Delta \Omega_n$ and $\Delta x_i $ is the distance between the points on the imaginary and the real axis and $ G_n=\sqrt{\Delta \Omega_n}G(i\Omega)$.
Note that the kernel for the analytic continuation has to be proportional to $K_{n,i} \propto \Delta x_i/(i\Omega_n - x_i)$, but it could be multiplied by an arbitrary separable weight function, which will only modify the metric in which the resulting singular functions are orthonormal. 

We have chosen a normalization such that the resulting imaginary axis singular-vectors will be automatically normalized using a standard metric $\int U_\alpha(i\Omega) U_{\alpha'}(i\Omega)d\Omega=\delta_{\alpha,\alpha'}$, as it will be shown below.
It is also important to recognize that the two meshes (on the real and the imaginary axes) are not equal. The real frequency mesh only needs to extend up to the selected high-energy cutoff (say $L$). This also required to be very precise in this interval with many points, as these functions strongly oscillate on the real axis. In contrast, the imaginary axis mesh needs to extend far beyond the scale of $L$. However as the functions are more smooth, a fewer points are typically needed. The rational for having a larger energy cutoff on the imaginary axis lies in the fact that any feature on the real axis, which is bounded in the interval $|x| < L$, will taper off slowly on the imaginary axis with a behavior like $1/(L^2+\Omega_n^2)$ for bosonic quantities. However, quantities on the imaginary axis are very smooth, and in particular, tails require a small number of points distributed in the logarithmic mesh.

It is obvious from Eqs.~\ref{Esvd1} and \ref{Ker:svd} that $\sqrt{\Delta \Omega_n} \; G(i\Omega_n) = \sum_{i} K_{n,i} A(x_i)$. 
Next, we perform the singular-value decomposition of the Kernel $K_{n,i}$ Eq.~\ref{Ker:svd}, and obtain $ K(\Omega_n,x_i) = \sum_\alpha u_\alpha (i\Omega_n) s_\alpha v_\alpha(x_i)$, where $s_\alpha$ are the singular values, and $u_\alpha(i\Omega_n)/\sqrt{\Delta \Omega_n}\equiv U_\alpha(i\Omega_n)$ are the desired SVD-basis functions. Now we see that
\begin{eqnarray}
\sum_n U_\alpha(i\Omega_n) U_{\alpha'}(i\Omega_n) \Delta \Omega_n = \sum_n u_\alpha(i\Omega_n) u_{\alpha'}(i\Omega_n) = \delta_{\alpha,\alpha'}
\nonumber 
\end{eqnarray}
because of the unitarity of the singular eigenvectors $u_{\alpha}(i\Omega_n)$, which proves that SVD-basis functions are an orthonormal basis. As it turns out, only a small number of singular values $s_\alpha$ are nonzero, because the Kernel for analytic continuation is known to be singular. Consequently this SVD-basis is the minimal orthonormal basis for representing Matsubara quantities.
We use a fine tangents mesh on the real frequency axis $x_i$,  and a different more coarse tangents mesh combined with logarithmic tails on the Matsubara axis for $\Omega_n$, and we can afford here a large number of real-frequency points (thousands) and also several hundred on the imaginary axis.

Next we represent the polarization in band basis $p_{ij}(\vk,\vq,i\Omega_n)$ in terms of these basis functions, i.e.
\begin{eqnarray}
p_{ij}(\vk,\vq,i\Omega_n) = \sum_\alpha U_\alpha(i\Omega_n) {\cal P}^\alpha_{ij}(\vk,\vq)
\end{eqnarray}
where ${\cal P}^\alpha_{ij}(\vk,\vq)$ are coefficients in this SVD-basis. The crucial point is that the number of coefficients $\alpha$ is much smaller than the number of needed Matsubara points. For example, to achieve the precision of polarization $p_{ij}(\vk,\vq,i\Omega_n)$ up $10^{-10}$, we typically need 30 coefficients. In this way, using these 30 coefficients on the SVD basis, we can then compute polarization on a much larger number of Matsubara frequencies.

First, we compute coefficients for polarization in band-basis ${\cal P}^\alpha_{ij}(\vk,\vq) = \sum_{n} \Delta\Omega_n U_\alpha(i\Omega_n) p_{ij}(\vk,\vq,i\Omega_n)$ inside the tetrahedron method so that we do not need to store large arrays $p_{ij}(\vk,\vq,i\Omega_n)$, and we rather store only the coefficients ${\cal P}^\alpha_{ij}(\vk,\vq)$. This is a simple matrix-matrix product and can be done very quickly, as there is a small number of basis functions $U_\alpha$. Next, we use these coefficients to get the dielectric matrix on a product basis in two steps:
\begin{eqnarray}
C^\alpha_{l,l'}(\vq) &=& \sum_{i,j,\vk}\widetilde{M}^*_{l,ij}(\vk,\vq)  {\cal P}^\alpha_{ij}(\vk,\vq) \widetilde{M}_{l',ij}(\vk,\vq)
\label{less_slow}\\
(1-\varepsilon)_{ll'} &=& \sum_\alpha U_\alpha(i\Omega) C^\alpha_{l,l'}(\vq) 
\label{much_less_slow}
\end{eqnarray}
The crucial point is that the product Eq.~\ref{less_slow} can be done faster than the product in the original Eq.~\ref{slowest}, when the number of 
coefficients ${\cal P}^\alpha$ is smaller than the number of Matsubara points $i\Omega_n$. There is some overhead due to the second step Eq.~\ref{much_less_slow}, but this is quite fast, because the product basis $l$ is much smaller in dimension than the square of the number of bands $i\otimes j$.

Finally, when comparing this SVD-basis implementation versus the convolution with spline interpolation, as explained in Eq.\ref{Eq:convol3}, we found somewhat mixed results. While both methods work well, the spline interpolation seems to be slightly more robust in the cases we tested.
This is because for Pade analytic continuation, we need to compute self-energy to extremely high precision, and consequently, we found that around 30 coefficients ${\cal P}^\alpha$ are necessary. However, $W(i\Omega_n)$ seems to be quite a smooth function of Matsubara frequency in all cases we tested, therefore with around 32-64 Matsubara points, we could also achieve extremely accurate spline for the screened interaction $W(i\Omega_n)$. Therefore in the test cases presented here, the SVD implementation did not significantly improve over the previously discussed spline interpolation. In cases with more nontrivial frequency dependence of $W(i\Omega)$, this SVD implementation will become more useful.

\subsection{Real frequency and quasiparticle band structure}
\label{RealAxis}

\subsubsection{Analytic continuation}
To obtain the self-energy on the real axis, we use the Pade analytic continuation method~\cite{Pade1,Pade0}, as well as the contour deformation technique, described below. The Pade method  is accurate at low to intermediate frequencies, when imaginary axis data is of very high accuracy.
We managed to arrange the numerics on imaginary axis to meet this goal and  to plot reliable band structures of materials around the Fermi level, as confirmed by the direct contour integration technique.
We emphasize that for metals, a few pole approximation in Pade-type fitting commonly employed in many other GW implementations, is usually not sufficient, and we avoid using such method. Instead we force the Pade approximate to go exactly through all Matsubara frequencies calculated (between 32-64), hence the number of poles in such an analytic function is large (between 30-62).
For future diagrammatic MC calculations, we anticipate using the maximum entropy method instead of Pade, as MC calculations are seldom of high enough precision to allow one to use Pade approximation.

\subsubsection{Contour deformation integration}
\label{Sec:contourDef}
We also implemented the contour deformation integration technique~\cite{contour0a,contour0b,contour1,contour2,contour3,PhysRevB.67.155208}, which is an alternative to the analytic continuation method and allows one to compute the self-energy directly on the real axis.
While this technique relies on a particular form of the $GW$ self-energy and is not straightforwardly  
extendable to higher-order Feynman diagrams, we want to point out that there is a recent promising progress in the direction of the higher-order evaluation of Feynman diagrams on the real-axis
using the algorithmic Matsubara integration~\cite{PhysRevB.99.035120,PhysRevB.102.045115}, whereby analytic expressions for higher-order Feynman diagrams are being derived, similar to contour deformation technique, for convolutions, and completely avoids integration over frequency. Note however that currently this has been applied only in the context of 
a single band Hubbard model, and the uniform electron gas~\cite{in_preparation}.

\begin{figure}
\begin{center}
\includegraphics[width=0.4\linewidth]{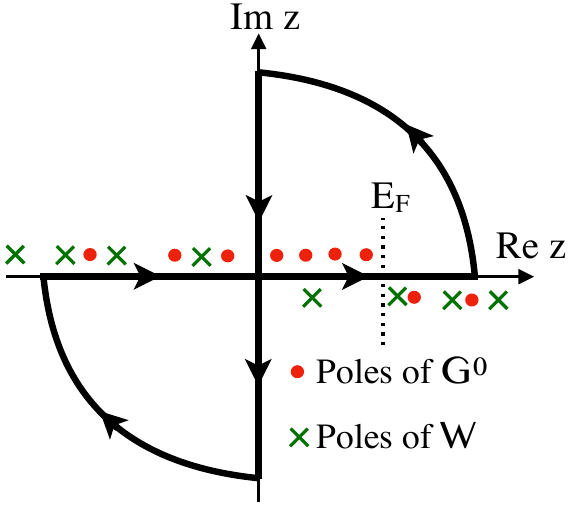}
\caption{ (Color online) 
Contour of the integration used to evaluate convolutions in GW approximation.
}
\label{fig-contour}
\end{center}
\end{figure}
The contour deformation is very successful in GW implementation because one needs to evaluate only simple integrals (convolutions) where all the poles of the integrand are either known exactly or can be avoided altogether by choosing the appropriate shape of the contour. For example, to evaluate the self-energy in Eq.~\ref{Sigma1}, one first takes the zero temperature limit, changing the sum over Matsubara frequencies into an integral, and one then uses the zero-temperature correlation functions $G^0$ and $W$, which are different from Matsubara and retarded analogs, and have the poles above (below) the real axis in the frequency below (above) $E_F$. The bosonic quantities, such as $W$, have a vanishing chemical potential, hence the poles jump across the real axis at the origin (see Fig.~\ref{fig-contour}).
The convolution Eq.~\ref{Sigma1} at zero-temperature takes the form
\begin{eqnarray}
\Sigma_\vk(\omega) =   -\int_{-\infty}^{\infty} \frac{dz}{2\pi i} W_\vq(z) G^0_{\vk-\vq}(\omega+z)
\label{RealSigma}
\end{eqnarray}
where we left out the matrix elements $M$ for simplicity and took into account that $W$ is even in frequency.
This convolution is actually carried out only for the correlation part of the self-energy, hence strictly speaking $W_\vq$ should be
understood as $W_\vq-V_\vq$ and $\Sigma_\vk$ should be understood as $\Sigma_\vk^c$. However, for simplicity, we keep here a simpler notation of $\Sigma_\vk$ and $W_\vq$.
When convoluting $G^0$ and $W$ we notice that one can choose a contour, depicted in Fig.~\ref{fig-contour}, which runs along the real axis from $-\infty$ to $\infty$, and it closes in such a way that one completely avoids the poles of $W$, and only poles of $G^0$ fall inside the contour. As a result, we do not need to know the residue of $W$ when carrying out the integral, and only the poles of $G^0$ and the residue at the poles are needed. These are particularly simple, namely, poles are at $z=\xi_{\vk-\vq,j}-\omega$, and residues are unity. We can  replace the integral over the real axis with the closed contour-integral over the shape depicted in Fig.~\ref{fig-contour}, minus the integral over the imaginary axis
\begin{eqnarray}
  \int_{-\infty}^{\infty} \frac{dz}{2\pi i} W_\vq(z) G^0_{\vk-\vq}(\omega+z)
  \nonumber\\
  =
  \oint \frac{dz}{2\pi i} W_\vq(z) G^0_{\vk-\vq}(\omega+z)
  \nonumber\\
  -\int_{+i\infty}^{-i\infty}\frac{dz}{2\pi i} W_\vq(z) G^0_{\vk-\vq}(\omega+z) 
\end{eqnarray}
The integral over the remaining semi-circles vanishes, because $G^0$ and $W^0$ fall off sufficiently fast, i.e., as $1/\omega$ and $1/\omega^2$.
The imaginary axis integral (the last term) is essentially the same integral with which we calculate the self-energy on the imaginary axis, and we know that the integrand is smooth and well-behaved, hence the spline integration discussed above gives very accurate results. To compute the self-energy on the real axis Eq.~\ref{RealSigma} we then just need to add the contour integral, which can be evaluated with the help of the residue theorem. The crucial point here is that the integrand is simple enough that we can analytically find all poles of the integrand inside the contour, and we can evaluate them. As discussed above, the poles of $W$ are all outside of our chosen contour, hence they do not contribute. The $G^0$ has poles at $z=\xi_{\vk-\vq,j}-\omega$ and for $z>0$ they appear in the first quadrant only.  The residue of $G^0$ in these poles is unity, hence the contour integral is $\sum_{\omega<\xi_{\vk-\vq,j}<E_F}W(\xi_{\vk-\vq,j}-\omega)$. On the other hand, when $z<0$ and $\xi_{\vk-\vq,j}<E_F$, the poles inside the contour appear in the third quadrant, and the integral is $-\sum_{E_F<\xi_{\vk-\vq,j}<\omega}W(\xi_{\vk-\vq,j}-\omega)$. The minus sign comes from the opposite orientation of the integral in the third quadrant. Putting all those terms together, we see that the self-energy on the real axis can be calculated in the following way
\begin{eqnarray}
  \Sigma_\vk(\omega) =   -\int_{-\infty}^{+\infty}\frac{dx}{2\pi } W_\vq(i x) G^0_{\vk-\vq}(\omega+i x) \nonumber\\
  -\sum_{\omega<\xi_{\vk-\vq,j}<E_F}W(\xi_{\vk-\vq,j}-\omega)\nonumber\\
  + \sum_{E_F<\xi_{\vk-\vq,j}<\omega}W(\xi_{\vk-\vq,j}-\omega)
\label{Eq:residue}  
\end{eqnarray}
%

While this integral appears almost as straightforward to implement as the imaginary axis self-energy
($ \Sigma_\vk(i\omega) =   -\int_{-\infty}^{+\infty}\frac{dx}{2\pi }  W_\vq(i x) G^0_{\vk-\vq}(i\omega+i x) $), the overhead in calculating $W_\vq$ (or dielectric matrix $\varepsilon$) at numerous additional points along the real axis incurs a significant computational overhead.
To evaluate the residues in Eq.~\ref{Eq:residue} we use the real frequency mesh with energy spacing of $10\,$mHa, which requires an additional 74 points on the real axis for a typical 10eV window of band structure plot. In addition, we use 32 points (or 64 points for checking the convergence) of non-uniformly distributed points along the imaginary frequency axis between 0 to 20*i mHa.

The comparison of Pade continuation with contour deformation integration is presented in Fig.~\ref{fig-matrix}. The difference is barely noticeable in the frequency range of interest. This is because the self-energy in these moderately correlated systems is relatively featureless. In $3d$ metallic systems the differences are larger, but this is left for future studies.

%
%
%

\subsubsection{Interpolation of band structure}
To obtain the band structure plots, we implemented two complementary techniques: the Wannierization using wannier90~\cite{PhysRevB.56.12847,RevModPhys.84.1419}, as well as interpolation using technique of Refs.~\cite{Pickett_method,Pickett_method0}. The two methods are compared in Figs.~\ref{fig-metal}. They give almost identical band structures when the number of momentum points in the calculation is large, for example, $16 \times 16 \times 16$ mesh. When the number of momentum points is small, for example $4\times 4 \times 4$, both band structures are relatively inaccurate, as the Fermi surface singularities are not properly resolved. We want to point out that this is very different from typical DFT calculation, where the convergence with the momentum space mesh is very rapid, as the semilocal correlations are quite insensitive to the quality of the momentum space mesh.

\noindent\textbf{Wannierization:}
The Wannierization requires two objects, the projection to local orbitals $A^\vk_{ij}=\braket{g_i|\psi_{\vk,j}}$ and the overlaps between Bloch orbitals at neighboring k-points $M^{\vk,\vb}_{i,j}=\braket{\psi_{\vk,i}|e^{-i\vb\vr}|\psi_{\vk+\vb,j}}$. Here $g_i$ is a chosen local orbital and $\psi_{i\vk}$ are Kohn-Sham bands. The latter is closely related to the overlap between the product basis and Kohn-Sham bands, i.e, the matrix elements $M_{\alpha,ij}(\vk,\vq)=\braket{\chi_{\alpha}^\vq|\psi_{\vk,i}\psi^*_{\vk-\vq,j}}$ defined above. Indeed, if we choose $\braket{\vr|\chi_\alpha^\vq} \equiv e^{i\vq\vr}$ in the muffin-thin sphere, and we choose the $\vG=0$ function in the interstitials, then $M^{\vk,\vb}_{i,j}=M^*_{\alpha,ij}(\vk,-\vb)$, hence these matrix elements are easily computed with existing GW machinery.

Within LAPW method, the overlaps $A^\vk_{ij}$ is readily  available for all functions in the muffin-thin sphere, including $u_l(r) Y_{lm}(\vr)$, $\dot{u}_l(r) Y_{lm}(\vr)$ and local orbitals $u^{LO}_{l}(r)Y_{lm}(\vr)$. We use singular value decomposition (SVD) to find the linear combination of local orbitals, which have the largest overlap for a certain set of bands that are the target of wannierization. More precisely, we first compute the overlaps
\begin{equation}
  \braket{u^\kappa_{lm} Y_{lm}|\psi_{\vk,j}} = A^\vk_{\kappa l m,j},
\end{equation}  
where $\kappa$ is a combined index for $u_l$, $\dot{u}_l$ and $u^{LO}_l$. Notice that in this step we orthogonalize $u^{LO}_l$ so that we have orthogonal basis $\braket{u^\kappa_{lm}|u^{\kappa'}_{l'm'}}=\delta_{\kappa l m,\kappa' l'm'}$.
Next we perform SVD on the local component 
\begin{equation}
\sum_{\vk} A^\vk_{\kappa l m,j} = U_{\kappa l m,i} s_i V^T_{i,j} ,
\end{equation}
where $s_i$ are the singular values.
If the number of targeted bands is $n$, we choose the largest $n$ singular values $s_i$, and create the linear combination of local orbitals with them
\begin{equation}
\braket{r|g_i} = \sum_{\kappa l m} U_{\kappa l m, i} u^\kappa_{l}(r) Y_{lm}(\vr)
\end{equation}
so that the local component of the needed overlaps are
\begin{equation}
\sum_\vk A_{ij}^\vk  = s_i V^T_{ij}
\end{equation}  
and are guaranteed to be non-vanishing. Of course matrix element $A_{ij}^\vk$ could still vanish at a particular momentum point, but on average it must be large, as we chose the largest $n$ eigenvalues $s_i$ in SVD decomposition. The above-defined quantities are finally used as input to the Wannier90 software. \\

\noindent\textbf{Band energy interpolation:}
 This technique is an alternative to the Wannierization technique (see Refs.~\cite{Pickett_method,Pickett_method0}) and relies on the fact that the quasiparticle energy is a scalar and hence invariant to all operations of the space group. The quasiparticle energy at each momentum point $\vk$ can be expanded as
\begin{eqnarray}
  \varepsilon(\vk) = \sum_m a_m S_m(\vk)
\label{Eq:epski}  
\end{eqnarray}
where $S_m(\vk)$ is the star of the lattice, i.e.,
\begin{equation}
S_m(\vk)   = \frac{1}{N_{sym}}\sum_{sym} e^{i\vk \Gamma_{sym} \vR_m}
\end{equation}  
and $\Gamma_{sym}$ are all symmetry operations of the lattice, and $\vR_m$ are the real space lattice vectors. Without loss of generality, we choose $m=0$ when  $\vR_m=0$. Notice that
$S_m(\vk)$ has the full symmetry of the crystal and is a scalar of the lattice space group. We should use here a considerably larger number of lattice vectors $\vR_m$ as compared to the number of simulated momentum points in the first Brillouin zone.

In this method, we require $\varepsilon(\vk)$ to coincide with the computed values of the quasiparticle dispersion ($\varepsilon_{\vk_i}$) on the discrete grid being used in the calculation, i.e., $\vk_i$ with $i={1,..,n}$ and at the same time is smooth throughout the Brillouin zone, which is achieved by a constrained minimization of the following functional:
\begin{eqnarray}
  {\cal R}\equiv \sum_{\vk_i} |\varepsilon(\vk_i)|^2 + c_1 |\nabla_\vk \varepsilon(\vk_i)|^2 + c_2|\nabla_\vk \varepsilon(\vk_i)|^4+\cdots
  \nonumber\\
  +\sum_i \lambda_i (\varepsilon(\vk_i) - \varepsilon_{\vk_i}),
\end{eqnarray}
where $\lambda_i$ are the Lagrange multipliers.
This functional can be rewritten in real space by the help of definition Eq.~\ref{Eq:epski}
\begin{eqnarray}
{\cal R} = \sum_m a_m^2 (1+ c_1 R_m^2 + c_2 R_m^4+\cdots)
  \label{Eq:sum1}\\
+\sum_i \lambda_i(\sum_m a_m S_m(\vk_i)-  \varepsilon_{\vk_i})
\label{Eq:sum2}
\end{eqnarray}
Here $c_i$ are some coefficients that regularize the dispersion, and we typically use $c_1=-2\times 0.25/R_{nn}^2$, $c_2 = (0.25)^2/R_{nn}^4$, and $c_3=0.25/R_{nn}^6$, where $R_{nn}$ is the nearest-neighbor distance so that the first part of the functional has a particularly simple form $a_m^2([1-0.25(R_m/R_{nn})^2]^2+0.25(R_m/R_{nn})^6)$.

Ref. ~\cite{Pickett_method} pointed out that $m=0$ term in Eq.~\ref{Eq:sum1} is harmful as it forces the average of the band to vanish, while from definition Eq.~\ref{Eq:epski} it follows that it should be equal to the center of the band, i.e.,
\begin{equation}
a_0 = \frac{1}{N_k}\sum_{i}   \varepsilon_{\vk_i}
\end{equation}  
therefore it is best to drop $m=0$ term in Eq.~\ref{Eq:sum1}  and
minimize
\begin{eqnarray}
{\cal R} = \sum_{m=1}^N a_m^2 (1+ c_1 R_m^2 + c_2 R_m^4+\cdots)
  \label{Eq:sum3}\\
+\sum_i \lambda_i(\sum_{m=0}^N a_m S_m(\vk_i)-  \varepsilon_{\vk_i})
\label{Eq:sum4}
\end{eqnarray}
Here $N$ has to be substantially larger that the number of points in the calculation, i.e., 
at least three to four times larger.

The constrained minimization can be performed analytically, and it requires only inversion of a matrix and matrix vector multiplication. For more detailed information of how to solve this minimization problem, the reader is referred to Ref.~\cite{Pickett_method}.

\subsubsection{The quasiparticle dispersion, scalar versus matrix form}
\label{sef:quasiparticle_dispersion}

We implemented the so-called G$_0$W$_0$ and GW$_0$ methods in both the scalar and the matrix form. In all cases, we compute screened interaction $W_0$ from Kohn-Sham Green's function $G_0$. In G$_0$W$_0$ we convolve $W_0$ with Kohn-Sham Green's function $G_0=1/(\omega+\mu-\varepsilon_\vk^0)$ using Eq.~\ref{Eq:convol3}. Here $\varepsilon_\vk^0$ is the Kohn-Sham energy.
In GW$_0$ method, the single-particle Green's function is determined self-consistently and is approximated with the quasiparticle form at every iteration.

\begin{figure}[bht]
\includegraphics[width=\linewidth]{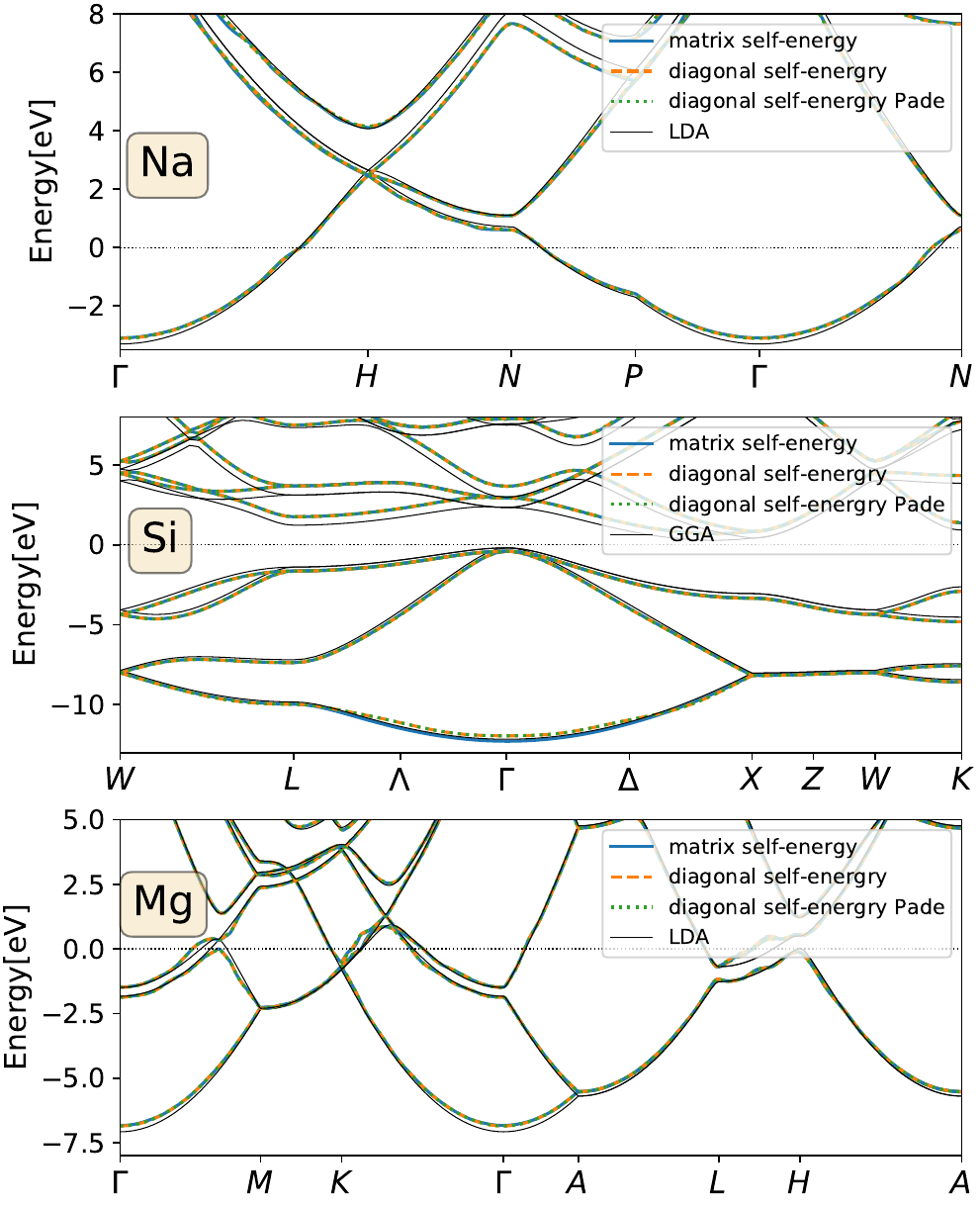}
\caption{ (Color online) Comparison of matrix self-energy to diagonal self-energy approximation in Na, Si and Mg using G$_0$W$_0$ approximation and contour deformation integration as well as Pade analytic continuation. Na and Mg band-structure is computed with $16\times 16\times 16$ k-point mesh and Si with  $6\times 6\times 6$ k-point mesh.
Interpolation is performed with a maximally localized wannier function algorithm.  
}
\label{fig-matrix}
\end{figure}

The scalar (non-matrix) approximation is most commonly used in GW, and its validity has been recently challenged in such a simple system as Si~\cite{PhysRevB.104.165111}. Therefore we checked the difference between the matrix form and the diagonal form of the self-energy for the systems we study here, including Si, Na, and Mg (see Fig.~\ref{fig-matrix}). We use the contour integration technique for both the matrix and diagonal self-energy and we also compared it with Pade continuation for diagonal self-energy. The calculation is converged with $6\times 6\times 6$ k-points for Si, and $16\times 16\times 16$ k-points for Na. Fig.~\ref{fig-matrix} shows that the diagonal self-energy approximation, as implemented here and explained below, shows almost no difference with the full matrix form, hence conventional scalar form is definitely justified at least for moderately correlated systems studied here. We checked that in more correlated $3d$ systems the matrix self-energy does make a difference, as the interaction in general increases hence GW bands become substantially different from DFT bands. It is also worth mentioning that Pade analytic continuation is excellent in these materials because the self-energy has very little frequency structure in the range of bands we are interested in.

In all cases, we are searching for the frequency $\omega$ where the interacting Green's function has poles, or equivalently, the zeros of the following matrix equation
\begin{eqnarray}
\omega I -\varepsilon_\vk^0  -\Sigma_\vk(\omega) + V_{xc} = 0
\end{eqnarray}
Here $\Sigma_\vk(\omega)=\Sigma^x_\vk + \Sigma^c_\vk(\omega)$ is the sum of exchange and correlation self-energy, and $\varepsilon_\vk^0$ is the diagonal Kohn-Sham energy in the Kohn-Sham band basis.

We use the linearized form of the self-energy to determine the poles of Green's function, i.e., we expand
\begin{equation}
\Sigma_\vk(\omega) = \Sigma(\varepsilon_{\vk,i}) + (I-Z_\vk^{-1}) (\omega-\varepsilon_{\vk,i})
\end{equation}  
with
$I-Z^{-1}_\vk = d\Sigma(\omega)/d\omega|_{\omega=\varepsilon_{\vk,i}}$ is the quasiparticle renormalization amplitude evaluated at the quasiparticle energy $\varepsilon_{\vk,i}$. This leads to the following eigenvalue problem
\begin{eqnarray}
Z_\vk^{-1} (\omega-\varepsilon_{\vk,i}) + (\varepsilon_{\vk,i} I-\varepsilon_\vk^0) - \Sigma_{\vk}(\varepsilon_{\vk,i})+V_{xc}=0
\end{eqnarray}  
or equivalently
\begin{eqnarray}
\omega=\varepsilon_{\vk,i} + Z_\vk^{1/2}\left(\varepsilon_\vk^0-\varepsilon_{\vk,i} I +\Sigma_{\vk}(\varepsilon_{\vk,i})-V_{xc}\right) Z_\vk^{1/2}
\label{Eq:QP1}
\end{eqnarray}
Since we are looking for the real solutions of this equation, we make all quantities in the above equation Hermitian, i.e.,
$\Sigma_\vk(\varepsilon_{\vk,i}) \leftarrow (\Sigma_\vk(\varepsilon_{\vk,i}) +\Sigma_\vk^\dagger(\varepsilon_{\vk,i}) )/2$.

Both G$_0$W$_0$ and GW$_0$ are traditionally solved in the scalar form, namely,
the self-energy and exchange correlation potential are approximated by the band-diagonal  form, i.e.,
$\Sigma_{\vk,i}(\omega) = \braket{\psi_{\vk,i}|\Sigma_\vk(\omega)|\psi_{\vk,i}}$, where $\psi_{i,\vk}$ are Kohn-Sham eigenvectors, hence
$Z_\vk$ are numbers, evaluated for each band $Z_{\vk,i}$ and the quasiparticle energies of band $i$ are
\begin{eqnarray}
  \omega^{qp}_i= \varepsilon_{\vk,i} +{ Z_{\vk,i}}{\left(\varepsilon^0_{\vk,i}-\varepsilon_{\vk,i} +\Sigma_{\vk}(\varepsilon_{\vk,i})-V_{xc}\right)}
\label{Eq:GW0}  
\end{eqnarray}

For the case of G$_0$W$_0$, the self-energy is computed by the Kohn-Sham band energies, $\varepsilon_{\vk,i}=\varepsilon^0_{\vk,i}$, hence self-energy can also be expand around Kohn-Sham energies, to get
\begin{eqnarray}
\omega_{G_0W_0,i}^{qp}= \varepsilon^0_{\vk,i} +Z_{\vk,i}{\left(\Sigma_{\vk}(\varepsilon^0_{\vk,i})-V_{xc}\right)}
\end{eqnarray}

In the case of GW$_0$ the self-energy is computed using the self-consistent quasiparticle energies $\varepsilon_{\vk,i} \leftarrow \omega^{qp}_i$ from previous iterations from the Eq.~\ref{Eq:GW0}, and the iterations are continued until $\varepsilon_{\vk,i} =\omega^{qp}_i$ up to some precision. 


Finally, when using the matrix form of the self-energy and the exchange correlation potential, we construct a Hermitian Hamiltonian from Eq.~\ref{Eq:QP1}
\begin{eqnarray}
H^{qp}_\vk = Z_\vk^{1/2}\left(\varepsilon_\vk^0-\varepsilon_{\vk,i} I +\Sigma_{\vk}(\varepsilon_{\vk,i})-V_{xc}\right) Z_\vk^{1/2}  
\end{eqnarray}
and solve for the eigenvalue $\lambda_i$, for which the eigenvector is the closest to unity eigenvector with component $i$ close to $1$, and zero otherwise. Clearly, we need to construct different Hamiltonian $H^{qp}_\vk$ for each band $i$, and take only one eigenvalue from the set of eigenvalues of this Hamiltonian. The quasiparticle energy is finally given by $\omega^{qp} = \varepsilon_{\vk,i} + \lambda_i$, as is clear from Eq.~\ref{Eq:QP1}. For G$_0$W$_0$ we can equate $\varepsilon_{\vk}$ with $\varepsilon^0_{\vk}$ in the above equation, which avoids the need for self-consistency. In GW$_0$ we require self-consistency in computing the self-energy, hence the expansion is also done around the current quasiparticle band energy.

When comparing the matrix form of the self-energy with the diagonal scalar approximation in Fig.~\ref{fig-matrix}  we notice that apart from a small downward shift of the first band in Si (around -12eV) there is no noticeable difference between the diagonal and matrix form of the self-energy. In particular, all metals studied here show no appreciable change when the off-diagonal self-energy is included. We notice that both the exchange self-energy and DFT semi-local exchange correlation potential are not very small, while the correlation self-energy tends to be somewhat smaller. However, their total effect is small as can be directly checked by evaluating the difference between the eigenvalue from Eq. ~\ref{Eq:QP1} and its diagonal equivalent Eq.~\ref{Eq:GW0}. This difference tends to be around mHa for relevant bands in the plot.

%

\subsection{Miscellaneous}

There are several important technical details of the implementation, which are not going to be extensively reviewed here, because they have been nicely explained in other works, for example in Ref.~\cite{Gap2} and Ref.~\cite{Bluegel}. 
\begin{itemize}
\item[a)] The algorithm to compute the matrix elements of the bare Coulomb repulsion has been thoroughly worked out in Ref.~\cite{Gap2}, and we followed their implementation. 

\item[b)] The inclusion of core states in the calculation is an important advantage of such an all-electron implementation. Here we again follow the implementation of Ref.~\cite{Gap2} and include core states in the basis. They contribute to the product basis, to the polarization calculation, and to the single-particle Green's function. 

\item[c)] We also implemented the $\vq\rightarrow 0$ limit as in Refs.~\cite{Gap2}, and with a few more tricks from Ref.~\cite{Bluegel,Friedrich_2012}. This analytic treatment of small $\vq$ limit of polarization and the Coulomb repulsion is essential, as the number of $\vq$-points is still quite limited, and we can not afford to drop $\vq=0$ point, rather we worked out the analytic limit of polarization using $\vk\cdot \vp$ perturbation theory. It turns out that $P_{0,0}(\vq\rightarrow 0)$, $P_{0,\vG}(\vq\rightarrow 0)$ and $P_{\vG,0}(\vq\rightarrow 0)$ are proportional to $\vq^2$, $\vq$, and $\vq$ respectively, so that even though the Coulomb repulsion is diverging at $\vq\rightarrow 0$, the dielectric constant is not, and its analytic treatment requires one to compute the matrix elements of the momentum $i\nabla$ operator, similarly as in the calculation of the optical conductivity. The term proportional to $q^2$ and $q$ is usually called head and wings, respectively.

When summing the terms that are divergent at $\vq\rightarrow 0$ but integrable, we have to add the correction due to a finite number of momentum points in the $\vq$ mesh. The divergent terms can have either $1/q^2$ or $1/q$ behavior, and would require one to sum
$\sum_\vq \frac{a_n}{|\vq+\vG|^n}$ where $n=1$ or $n=2$.
We first evaluate the sum by dropping the divergent term $\vq=\vG=0$, and later we add the correction $\Delta_c$, which vanishes for very dense momentum mesh, but gives correction when momentum mesh is sparse. Specifically,
\begin{eqnarray}
\sum_{\vq,\vG}\frac{a_n}{|\vq+\vG|^n}\rightarrow \sum'_{(\vq,\vG)\ne 0}\frac{a_n}{|\vq+\vG|^n}+a_n\; \Delta^n_C
\end{eqnarray}
with
\begin{eqnarray}
  \Delta^n_c = \sum_\vG\frac{V}{(2\pi)^3}\int d^3\vq \frac{e^{-\alpha(\vq+\vG)^2}}{|\vq+\vG|^n}
  \nonumber\\
  - \sum'_{(\vq,\vG)\ne 0}\frac{e^{-\alpha(\vq+\vG)^2}}{|\vq+\vG|^n}
  \label{Eq:DEL}
\end{eqnarray}
The first term in Eq. ~\ref{Eq:DEL} is evaluated analytically, while the second term is evaluated on the discrete mesh.
Here we added a small positive constant $\alpha$ in the exponent, which does not change the nature of the $\vq\rightarrow 0$ divergency, but makes the integral converge fast.
If the $\vq$ mesh is dense, $\Delta_c^n$ vanishes, while a sparse $\vq$ mesh has mostly contribution at small $\vq+\vG$ and is hence very weakly dependent on $\alpha$ for small $\alpha$.

\item[d)]
In contrast to insulators, the metallic systems also contain the so-called Drude term as part of the dielectric matrix. This is in addition to other singular terms arising in insulators, which were briefly discussed above. Here we show where Drude's term comes from, and how we treat it.
In the eigenbasis of the Coulomb repulsion, we know that the singular eigenvalue in the limit $\vq\rightarrow 0$ is $4\pi/q^2$, and the exact eigenvector is $e^{i\vq \vr}/\sqrt{V_{cell}}$. This is because the Coulomb repulsion in the plane wave basis and in the eigenbasis of the Coulomb matrix (expressed in terms of LAPW product functions) are similar matrices, and its non-degenerate singular part is therefore unique. The projection to the Kohn-Sham bands of this singular eigenvector therefore is
\begin{eqnarray}
  \widetilde{M}(\vk,\vq\rightarrow 0)_{l=0,ij} \approx \sqrt{\frac{4\pi}{\vq^2}}\braket{\frac{e^{i\vq\vr}}{\sqrt{V_{cell}}}|\psi_{\vk,i}\psi^*_{\vk+\vq,j}}
\nonumber\\
  \rightarrow \delta_{i,j} \sqrt{\frac{4\pi}{\vq^2\, V_{cell}}}
\end{eqnarray}
Next, we want to evaluate the dielectric function in the same $\vq\rightarrow 0$ limit, which follows from Eq.~\ref{Eqeps}
\begin{eqnarray}
  1-\varepsilon_{l=0,l'=0}\approx \frac{4\pi N_s}{\vq^2 V_{cell}}\sum_{\vk,i}\frac{
  \left(-\frac{d f}{d\varepsilon}(\xi_{\vk,i})
  \right)(\varepsilon_{\vk+\vq,i}-\varepsilon_{\vk,i})}
  {i\Omega_n-(\varepsilon_{\vk+\vq,i}-\varepsilon_{\vk,i})) }
\label{limit1}
\end{eqnarray}
where $\xi_{\vk,i}=\varepsilon_{\vk,i}-\mu$.
Within $k\cdot p$ perturbation theory, the difference of Kohn-Sham energies is
$\varepsilon_{\vk+\vq,i}-\varepsilon_{\vk,i}\approx \frac{\vq}{m}\braket{\psi_{\vk,i}|-i\nabla|\psi_{\vk,i}} \equiv \frac{\vq\cdot \vp^\vk_{i i}}{m}$
Inserting this expression into Eq.~\ref{limit1}, and expanding for small $\vq$, we get the following result
\begin{eqnarray}
  1-\varepsilon_{0,0}\approx \frac{4\pi N_s}{ (i\Omega_n)^2 V_{cell}} \sum_{\vk,i}
  {\left(-\frac{d f}{d\varepsilon}(\xi_{\vk,i})\right)\left( \frac{ e_\vq\cdot \vp^\vk_{i i}}{m}\right)^2 }
\label{limit2}
\end{eqnarray}
Note that we dropped the linear term because its contribution vanishes as it is odd in $\vq$.
Now we recognize the plasma frequency
\begin{eqnarray}
\omega_p^2 \equiv \frac{4\pi N_{s}}{m^2 V_{cell}} \sum_{\vk,i}
  {\left(-\frac{d f}{d\varepsilon}(\xi_{\vk,i})\right)\left( { e_\vq\cdot \vp^\vk_{i i}}\right)^2 }
\end{eqnarray}
in terms of which the Drude part of the dielectric constant is
\begin{eqnarray}
  1-\varepsilon_{0,0}(\vq\rightarrow 0)\approx \frac{\omega_p^2}{(i\Omega_n)^2}
\end{eqnarray}
Note that we need $\varepsilon^{-1}-1$ to compute the self-energy in Eq.~\ref{Eq:convol}. While this Drude term gives singular contribution on the real axis, it is however well behaved on the imaginary axis, as it takes the form $\varepsilon^{-1}-1 = \frac{1}{ 1+\omega_p^2/\Omega_n^2}-1$.

This Drude term, which appears at  $\vq=0$, is of course missed in the discrete sum of Eq.~\ref{Eqeps}, hence we need to add it to the so-called head part of the dielectric matrix, i.e.,
$\varepsilon_{0,0} \rightarrow \varepsilon_{0,0} - \frac{\omega_p^2}{(i\Omega_n)^2}$, before we invert the dielectric matrix  to compute $\varepsilon^{-1}$.

\end{itemize}

\section{Results}
\label{Results}

\subsection{Benchmarking and Validation in Insulators}


First, we describe our results for insulating systems to benchmark our GW implementation. We have computed GW quasiparticle energies and  band gaps for a set of prototypical insulating materials, such as Si, BN, SiC, MgO, ZnS, CdS, LiF, etc. The experimental band-gap ranges in this set of compounds between 1.2 and 14.2 eV. The 8$\times$ 8 $\times$ 8 k-point grid is considered here. The starting point for GW calculation in insulators is obtained from DFT-GGA simulation using PBE functional. It is worth mentioning here that the gap size does depend on the choice of the DFT exchange-correlation functional. However, the future diagrammatic Monte Carlo method, which sums up all relevant higher-order Feynman diagrams, should not anymore depend on the starting point as the higher-order counter-terms can be properly subtracted~\cite{Kun}.

\begin{table*}
\setlength{\tabcolsep}{1.0pt} 
 \begin{minipage}{1.0\linewidth}
\begin{center}
\begin{tabular}{|c|c | c | c |c| c| c|c | c|c|c|}
\hline
{Compound} & {a } & {PBE} & &{G$_0$W$_0$} &{G$_0$W$_0$} &{G$_0$W$_0$ } & {GW$_0$} &{GW$_0$}  &{GW$_0$} &  {Expt} \\
&(\AA)&&&(present)&(LAPW~\cite{Blaha_HLO})&(PAW~\cite{PhysRevB.75.235102})&(present)&(LAPW~\cite{Blaha_HLO})&(PAW~\cite{PhysRevB.75.235102})   & \\
\hline
\multirow{1}{*}{Si} & 5.430 ~\cite{PhysRevB.69.075102} & 0.573 & LO=0 & 1.095 & 1.03 & & 1.13 & 1.09 &&  \\
& & &LO=5   & 1.11 & 1.12 & 1.12& 1.19 & 1.19 &1.20 & 1.17\\
\hline
\multirow{1}{*}{BN} &3.615  & 4.472& LO=0&  5.97 & 6.04 & & 6.19 & 6.27 & &  \\
&&&LO=5  & 6.15 & 6.36 & 6.10& 6.39 & 6.61 &6.35& 6.1-6.4\\
\hline
\multirow{1}{*}{CdS} &5.832~\cite{CdS} &1.13& LO=0 & 1.88 & 2.02 & & 2.01 & 2.18 && \\
&&&LO=5   & 1.92 & 2.19 & 2.06& 2.05 & 2.38 &2.26&2.42   \\
\hline
\multirow{1}{*}{MgO} &4.213~\cite{MgO}  & 4.74 & LO=0& 7.04 & 7.08 &  & 7.45 & 7.52 && \\
&&&LO=5   & 7.22 & 7.52 & 7.25& 7.63 & 8.01 &7.72& 7.83 \\
\hline
\multirow{1}{*}{SiC}&4.358~\cite{SiC} & 1.36 & LO=0 & 2.13& 2.23 & & 2.25 & 2.36 & & \\
&&&LO=5   & 2.16 & 2.38 & 2.27& 2.27 & 2.53 &2.43&2.40   \\
\hline
\multirow{1}{*}{ZnS}&5.41 ~\cite{ZnS} &2.08 & LO=0& 3.19 & 3.15 &  & 3.44 & 3.35 & & \\
&&&LO=5   & 3.27 & 3.35 & 3.29& 3.48 & 3.61 &3.54&3.91 \\
\hline
\multirow{1}{*}{LiF} &4.028 & 9.08& LO=0& 12.96 & 12.36 &  & 13.45 & 13.98 & & \\
&&&LO=5   &  13.42 & 14.27 & 13.27& 14.18 & 15.13 & 13.96& 14.2  \\
\hline
\end{tabular}
\caption{Bandgap (in eV) of various insulators as computed in PBE and G$_0$W$_0$ approaches and their comparison with experiments and previous GW results using PAW and LAPW basis, which are quoted from Ref ~\cite{PhysRevB.75.235102} and Ref~\cite{Blaha_HLO} respectively. Comparisons of band-gap without LO and LO=5 are shown on the top and bottom respectively.  }
\label{tab:insu}
\end{center}
\end{minipage}
\end{table*}

\begin{figure*}
\includegraphics[width=\linewidth]{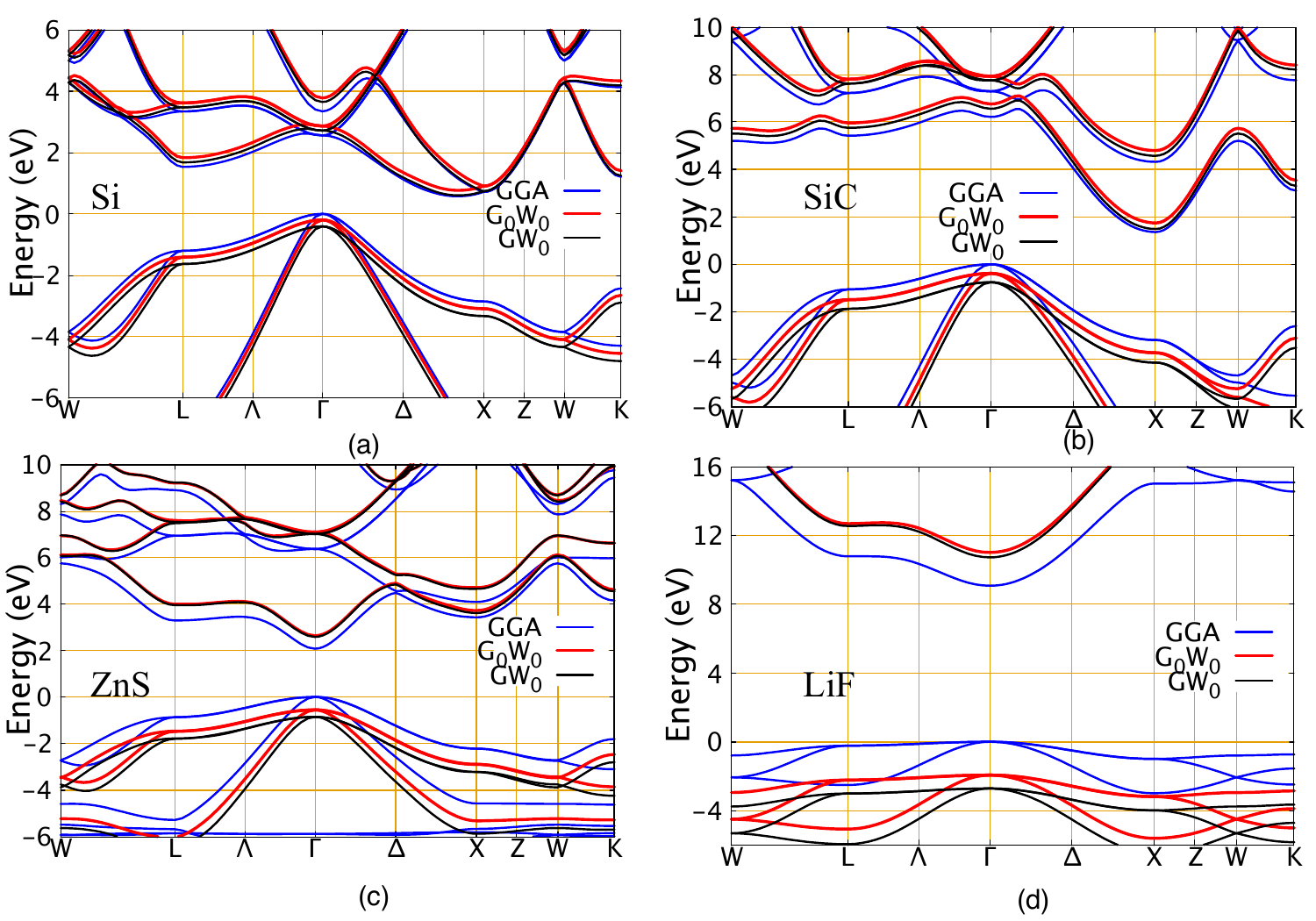}
\caption{ (Color online) Band structure for insulators as computed in GGA (PBE), G$_0$W$_0$, and GW$_0$ approximations without LOs for: (a) Si, (b)SiC, (c) ZnS, and (d) LiF. For each compound we notice an increase in the band-gap in either G$_0$W$_0$ or GW$_0$.  }
\label{fig-insulator}
\end{figure*}

The computed bandgap within G$_0$W$_0$ and GW$_0$ are summarized in Table~\ref{tab:insu}. We compare them with PAW~\cite{PhysRevB.75.235102} and previous LAPW~\cite{Blaha_HLO} results with and without additional local orbitals. We used the experimental lattice constants from the literature (see column 2), which are close to the values quoted by Shishkin and Kresse in Ref.~\cite{PhysRevB.75.235102}, but somewhat different than those used in Ref.~\cite{Blaha_HLO}. Experimental band gaps are quoted from Ref.~\cite{PhysRevB.75.235102}, which compares well with our results and previous literature. We also compare our results obtained with and without considering LOs, and as can be seen from the table, additional LOs typically increase the size of the gap. The energy levels for LOs are obtained from Ref.~\cite{Blaha_HLO}. 

As is well known, for every band-insulating compound, the bandgap increases in G$_0$W$_0$ as compared to DFT-PBE value. In GW$_0$ the band gap is further increased compared to G$_0$W$_0$, especially for wide gap insulators like LiF. 
We mention in passing that GW$_0$ calculation is a very cheap post-processing step, once G$_0$W$_0$ calculation is finished. This is because most of the computational time is spent in evaluating the screened interaction $W_0$, and once this is available, only the convolution Eq.~\ref{Eq:convol3} needs to be repeated several times to determine the self-consistent quasi-particle energies $\varepsilon_\vk$ from Eq.~\ref{Eq:GW0}.

As is clear from Table~\ref{tab:insu}, our results agree well with previous LAPW work by Jiang and Blaha ~\cite{Blaha_HLO}, and are also very close also to PAW results of Ref.~\cite{PhysRevB.75.235102}. 
The slight difference in the size of the band gaps between our results and those of Jiang and Blaha ~\cite{Blaha_HLO} is mainly due to the difference in the lattice constants used in the two calculations. For example, the computed band-gap of CdS using our code is 1.88 and 2.01 eV in G$_0$W$_0$ and GW$_0$, respectively, while it is 2.02 and 2.18 in Ref.~\cite{Blaha_HLO}. If we rerun Gap2 code (used in Ref.~\cite{Blaha_HLO}) on the experimental lattice constant quoted here, the band-gap is very close to our values, namely,  1.90 and 2.04 eV in G$_0$W$_0$ and GW$_0$ respectively. We also noticed in passing that somewhat smaller muffin-tin radii in combination with a bit larger plane wave cutoff (``RKmax'') tends to slightly increase the gap (within a percent) in most of the insulators. In our calculations we have not fine-tuned these values.

\begin{table*}
\setlength{\tabcolsep}{1.0pt} 
\begin{center}
\begin{tabular}{|l|c|c|c|c|c|c|}
\hline
Setup & gap G0W0 & $\Gamma-X$ gap G0W0 & gap GW0 & $\Gamma-X$ gap GW0 & P.B. size & eigen. size\\
\hline
 0 LO's, $L_{max}=6$, $PB_{emax}=20$H & 1.063 eV & 1.201 eV & 1.128 eV & 1.267 eV & 437 & 405\\
 5 LO's, $L_{max}=6$,  $PB_{emax}=20$H & 1.090 eV & 1.224 eV & 1.158 eV & 1.292 eV & 575 & 466\\
 5 LO's, $L_{max}=6$,  $PB_{emax}=\infty$ & 1.090 eV & 1.224 eV & 1.158 eV & 1.292 eV & 1407 & 544\\
 5 LO's, $L_{max}=10$, $PB_{emax}=20$H& 1.095 eV & 1.227 eV & 1.162 eV & 1.295 eV & 1013 & 820\\
 5 LO's, $L_{max}=10$, $PB_{emax}=\infty$& 1.095 eV & 1.227 eV & 1.162 eV & 1.295 eV & 2019 & 958\\
\hline
 Ref.~\cite{Blaha_HLO}, $a=10.23543\,a_B$ & 1.12 eV & & 1.19 eV & & &\\
 Ref.~\cite{Bluegel}, $a=10.26253\,a_B$  & 1.11 eV & & & & &\\
 Experiment & 1.17 eV & 1.25 eV & 1.17 eV & 1.25 eV & &\\
\hline
\end{tabular}
\end{center}    
\caption{
Convergence of gaps for Si with experimental lattice constant $a=10.262536\,a_B$, the plane wave cutoff for interstitial basis $RKmax=8$, and 
number of momentum points $4\times 4\times 4$. Here LO stands for the number of local orbitals. We choose the same local orbital energies as in Ref.~\cite{Blaha_HLO}. $L_{max}$ is the maximum orbital momentum $L$ allowed in the product basis and introduced above Eq.~\ref{BasisDiagonal2}. $PB_{emax}$ is the cutoff energy in Hartee's for including an orbital in product basis. P.B. size is the size of the product basis, namely the dimension of the index $\alpha$ in $M_{\alpha,ij}(\vk,\vq)$. eigen. size is the size of the eigenbasis of the Coulomb repulsion, i.e., the dimension of the index $l$ in $\widetilde{M}_{l,ij}(\vq,\vq)$.
}
\label{table:convergenceSi}
\end{table*}
In Table~\ref{table:convergenceSi} we show how the size of the gap depends on the parameters of the product LAPW basis. Here we use a converged number of Matsubara points (32 for evaluating $W_0$ and 160 for the convolution of $W_0$ and $G$). The important parameters are: the number of local orbitals (LO), the highest allowed orbital momentum of the product basis $L_{max}$ defined just above Eq.~\ref{BasisDiagonal2}, the maximum energy of the radial orbital included in the product basis $PB_{emax}$. Namely, when constructing the product basis, we always include all the basis-functions corresponding to occupied states as well as core state, however, we can neglect some radial basis functions, which are solutions of the Schroedinger equation at very high energy (beyond $PB_{emax}$). We start convergence tests with the cutoff $L_{max}=6$ and $PB_{emax}=20$Hartree above the Fermi energy, which gives a gap in Si within 3\% of the converged value. This requires the product basis size of 437, and the Coulomb eigenbasis size of 405. Clearly, in such an economic setup almost all basis functions are important, and hence calculation in eigenbasis does not speed up the calculation much.

Next, we add five LO's at the energies tabulated in Ref.~\cite{Blaha_HLO}, which converges the gap within 0.5\%, and increases the product basis for additional 138 functions, while the eigenbasis size is increased for only 61 functions.
Increasing $PB_{emax}$ to infinity changes the gap size for less than 0.2\%, however, it increases the product basis substantially to the size of 1407, i.e., additional 832 basis functions. Here the power of the Coulomb eigenbasis becomes apparent, as that basis increases for only 78 additional functions, i.e., one order of magnitude less than the number of functions added to the product basis. Finally, increasing $L_{max}$ from 6 to 10 adds an additional 0.5\% to the gap size, and increases the product basis for additional 438 functions, while the Coulomb eigenbasis is increased for 354 functions. Finally, increasing $PB_{emax}$ at already converged $L_{max}=10$ does not change the gap but increases the product basis substantially. Fortunately, the eigenbasis is increased much less. Hence the energy cutoff $PB_{emax}=20$Hartree (default in Gap2 code) allows one to substantially reduce the computational cost and reduce the product basis size and not affect the results much. At the same time, the Coulomb eigenbasis is a much more economic basis than the product basis to perform calculations of polarization matrix and $W$ matrix.

Finally, in Fig.~6 we plot the band structure along the high symmetry lines for selected insulators, namely Si, SiC, ZnS, and LiF. As is well known, the major effect of G$_0$W$_0$ and GW$_0$ as compared to DFT is the shifting of the valence and conduction bands away from each other to increase the gap size. The connectivity of the bands and the overall band-structure is only moderately changed from its DFT structure, and the band renormalization is also quite weak in most band insulators, except for LiF, where the band renormalization from GGA is quite strong. We notice that the shift in valence and conduction band in GW is present throughout the BZ and is not particular to a specific symmetry point.


\subsection{Results for Metallic Systems with Convergence Tests}
\label{Metals}
Many widely available software packages now support GW corrections to gaps in semiconductors, however, very few support GW calculation in the metallic system, and even fewer allow one to plot the band structure throughout the Brillouin zone.
%
%
This is due to the numerical difficulty in treating the Fermi surface singularity in metallic systems, which oftentimes leads to less accurate results on the Matsubara axis, and consequently extremely difficult analytic continuation to real frequency. 
Here we have improved the stability of the tetrahedron method, as implemented in Gap2 code~\cite{Gap2_code}, and improved the convolution between the $G$ and $W$, so that the standard Pade approximation is stable. 


\begin{figure}
\includegraphics[width=0.99\linewidth]{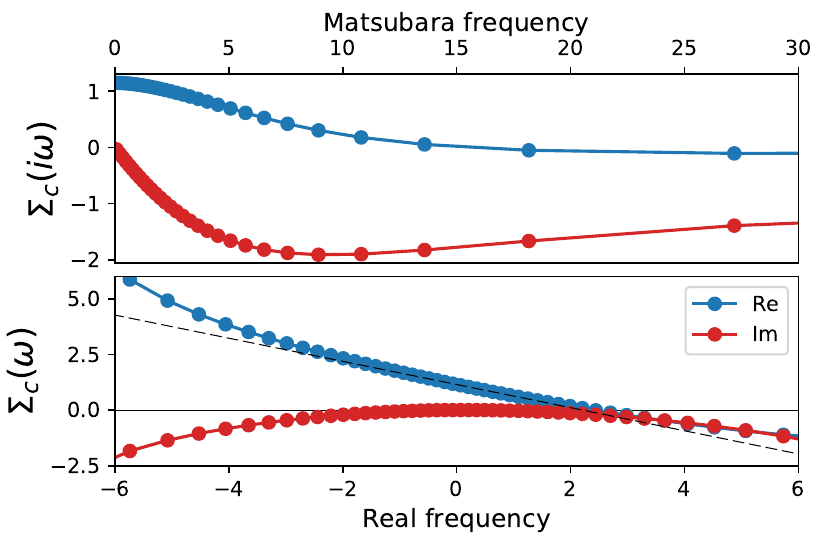}
\caption{(Color online) Self-energy  on Matsubara and real axis for Na at the $\Gamma$ point of the valence band. The straight line on real axis shows the quasiparticle approximation expanding around zero frequency.
}
\label{fig-sigma}
\end{figure}
In Fig.~\ref{fig-sigma} we show the correlation self-energy on the Matsubara and on the real axis for Na at $\Gamma$ point of the last valence band. The imaginary part of the self-energy on the real axis is roughly quadratic with a very large coherence scale, which is roughly proportional to the width of the parabola. The real part is linear at low frequency, however, at the frequency of the quasiparticle peak (around $-3\,$eV), $Re\Sigma$ substantially deviates from the straight line, hence simple quasiparticle approximation, which expands around zero frequency, would lead to smaller self-energy at $-3eV$, and consequently to larger bandwidth of Na. This demonstrates that accurate analytic continuation is crucial for extracting precise bandwidth of metals, as very accurate self-energy at finite frequency is required, beyond linear approximation. We also checked the precision of the Pade analytic continuation by comparing it to contour deformation integration in Fig.~\ref{fig-matrix}, which shows an excellent agreement between the two methods.

\begin{figure*}
\includegraphics[width=\linewidth]{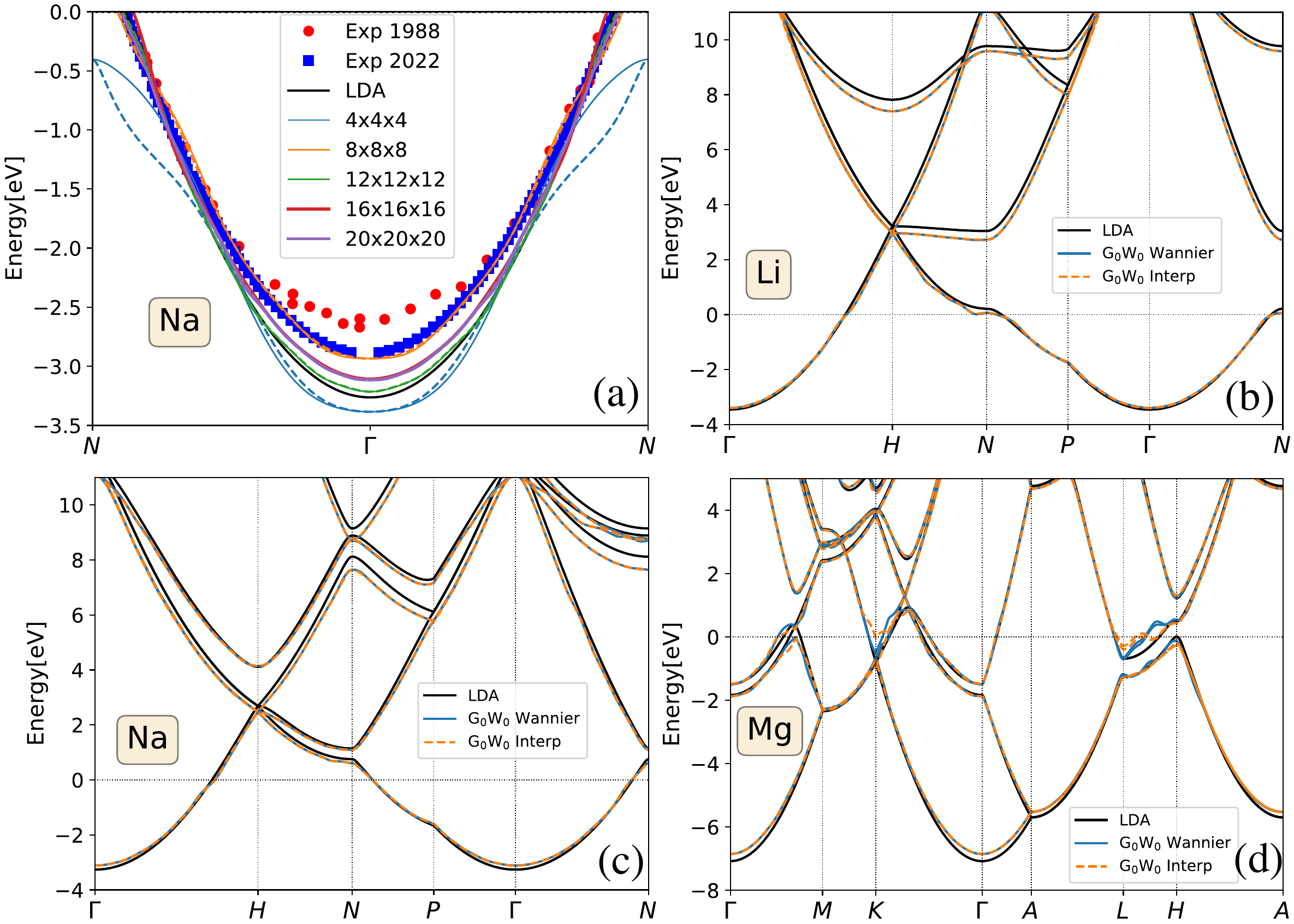}
\caption{ (Color online) (a) Convergence of the band-structure with
  momentum grid in Na, showing the bandwidth of the occupied bands for
  metals in GW approximation.  The dashed curves show the
  interpolation using method of
  Refs.~\cite{Pickett_method,Pickett_method0}, while the
  continous curves correspond to maximally-localized wannier functions
  interpolation~\cite{PhysRevB.56.12847,RevModPhys.84.1419}.  Note that the $8\times 8\times 8$ result is not yet converged, but is accidentally close to the newest
  experimental ARPES. The $16\times 16\times 16$ and $20\times 20\times 20$ curves are
  indistinguishable in this plot, hence converged.  Red dots (Exp 1988) and blue squares (Exp 2022) are the
  experimental ARPES data, which are
  reproduced from Refs.~\cite{PhysRevLett.60.1558} and ~\cite{Na_ARPES_new}, respectively. Band structure for
  elemental metals for (b) Li, (c) Na, and (d) Mg as computed in LDA
  and G$_0$W$_0$ at $16\times 16\times 16$ momentum mesh. The solid and the dashed line correspond to Wannier interpolation and the interpolation of  Refs.~\cite{Pickett_method,Pickett_method0} }
\label{fig-metal}
\end{figure*}
A somewhat surprising fact is that even though we use tetrahedron analytic integration over momentum points, we still find that a very large number of momentum points are necessary for converged results in metals. While even $4\times 4 \times 4$ grid gives approximate spectra which resemble LDA bands, the convergence with increasing momentum points is slow, and is presented in Fig~\ref{fig-metal} (a). For comparison, we also plot LDA values and ARPES data which are reproduced from Ref~\cite{PhysRevLett.60.1558} and Ref.~\cite{Na_ARPES_new}.
We notice that $4\times 4\times 4$ mesh does not have a Fermi surface crossing between $\Gamma-N$, hence the topology of the Fermi surface is wrong at this approximate mesh.
Moreover, the maximally localized wannier interpolation (dotted lines) is quite different from the interpolation of Ref.~\cite{Pickett_method} (straight line), which agree only in discrete points at $\Gamma$, $N$ and halfway between $\Gamma-N$, i.e., the points being used in the calculation. The bandwidth is severely overestimated, beyond LDA bandwidth. With $8\times 8\times 8$ mesh the Fermi surface and the bandwidth are accidentally very close to the experimental data of Ref.~\cite{Na_ARPES_new}. However, this is not a converged result within GW approximation, as $12\times 12\times 12$ mesh shows substantially larger bandwidth, close to LDA results. Only the $16\times 16\times 16$ and $20\times 20\times 20$ mesh agree, and can be taken as the converged result with GW approximation. The Na bandwidth within G$_0$W$_0$ is 3.12eV as compared to LDA value of 3.3eV, and ARPES results from 1988~\cite{PhysRevLett.60.1558}  of 2.65, and newer 2022 results~\cite{Na_ARPES_new} of 2.88 eV. We notice that the new ARPES bandwidth is much closer to GW prediction than the older results, but is still around 8\% too large. It is likely that this relatively moderate error will be eliminated by the proper inclusion of vertex corrections. We notice in passing that the inclusion of local vertex corrections, as implemented in DMFT, indeed agrees with the new ARPES rather well, with predicted bandwidth of 2.84eV~\cite{SubhasishMetals}.


In Fig.~\ref{fig-metal} we show band structure plots along high-symmetry lines for Li, Na, and Mg, and we present the bandwidth (energy difference between the $\Gamma$ point energy and the Fermi energy) in table III. We compare our results to those of Ref.~\cite{Kutepov_method}, and to the experiment.
First, we notice that the band structures of all these compounds are remarkably similar to the LDA (or GGA) results. As the Fermi surface is almost exactly spherical in these compounds, and the band structure is close to a renormalized free-electron solution in the proper periodic potential, the only relevant number in such calculations is the bandwidth.
We notice that the bandwidth is reduced as compared to LDA in all the compounds studied here. The range of band narrowing compared to LDA is about $\sim$ 2-7 \%, which is far smaller than in the experiment or reported in Ref.~\cite{Hybertsen_Na}. It is however quite similar to recently reported self-consistent quasi-particle GW  values in Ref.~\cite{Kutepov_method}. We also notice that our G$_0$W$_0$ results compare slightly more favorably with the experiment than the self-consistent quasi-particle GW  method, nevertheless, there is a substantial renormalization effect missing within G$_0$W$_0$ or QSGW method. These results, therefore, suggest that the vertex corrections beyond GW might be substantial even in these systems with predominantly $s$ and $p$ electrons. Such selected vertex corrections were studied in Ref.~\cite{Kutepov_method}, and with more phenomenological ansatz also in Ref.~\cite{SLouieSpinF}. The local vertex corrections were studied in Ref.~\cite{SubhasishMetals}, which predict bandwidth very close to the newer ARPES results~\cite{Na_ARPES_new}.
However, we believe that a more systematic approach offered by the diagrammatic Monte Carlo method would be very useful here, to understand the rate of the perturbation theory convergence with the perturbation order in metals with predominantly $s$ and $p$ electrons.

\begin{table*}
\begin{center}
\begin{tabular}{| c | c | c |c| c|}
\hline
{Compound} & {LDA}  &{G$_0$W$_0$ (present)}  & {Expt} &{QSGW~\cite{Kutepov_method}} \\
\hline
\multirow{1}{*}{Li} & 3.46 & 3.39 &  &\\

\hline
\multirow{1}{*}{Na}  & 3.30 & 3.12 & 2.65 ~\cite{PhysRevLett.60.1558}, 2.88~\cite{Na_ARPES_new} & 3.17 \\
\hline
\multirow{1}{*}{K} & 2.15 &2.00 & 1.6 ~\cite{PhysRevB.41.8075} & 2.07\\
\hline

\multirow{1}{*}{Mg} & 1.31,1.65, 6.89 & 1.29, 1.68, 6.66& 0.9, 1.7, 6.15 ~\cite{PhysRevB.33.3644} & \\
\hline
\end{tabular}
\caption{Bandwidth of occupied bands for elemental metals as computed in LDA and G$_0$W$_0$ approaches and their comparison with experiments and self-consistent quasi-particle GW (QSGW) which are adopted from Ref ~\cite{Kutepov_method}}
\label{tab:metal}
\end{center}
\end{table*}

\subsection{Scaling and Computational Cost}
One of the biggest bottlenecks in GW calculations is the computational cost of simulations and the scaling of the software. Although, theoretically GW scales as O(N$^4$), where $N$ is the number of bands, while DFT scales O(N$^3$), practically we find GW method is around two orders of magnitude slower compared to DFT even for the smallest single atom unit cell with only around hundred of bands~\cite{TMO1-SM}, and becomes even slower with increasing system size. Hence the search for greater efficiency of the GW implementation and GW algorithm has became one of the important research directions in the community~\cite{SM-GW,rocca_ab_2010,giustino_gw_2010,umari_gw_2010,govoni_large_2015,bruneval_accurate_2008,berger_ab_2010,gao_speeding_2016,liu_cubic_2016,foerster_on3_2011}. One possibility is to reduce the number of necessary unoccupied states and consequently reduce the scaling from O(N$^4$) to O(N$^3$)~\cite{liu_cubic_2016,govoni_large_2015,PhysRevB.101.035139}. Here we focus on the alternative direction in which we reduce the prefactor, and keep the O(N$^4$) scaling. This is because for higher-order Feynman diagrams, for which this software will be used, such a trick of reduced scaling is unlikely to be found. Hence, we here concentrate on optimizing the standard GW algorithm described in previous sections. 

\begin{figure}[bht]
\includegraphics[width=0.95\linewidth]{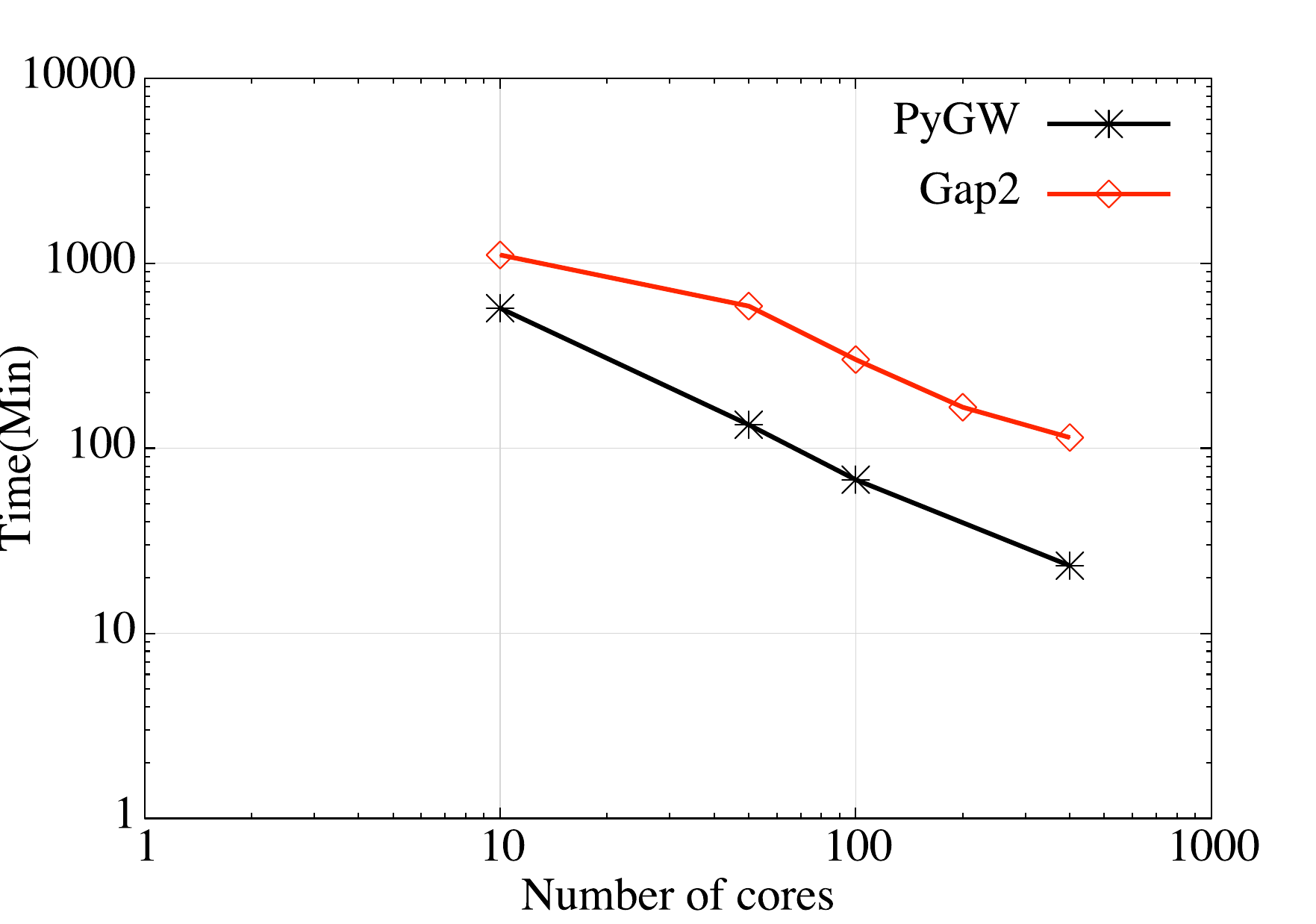}
\caption{ (Color online)Comparison of computational cost in PyGW and Gap2 codes: Logarithmic plot for computational cost in G$_0$W$_0$ calculation for MgO using PyGW and Gap2 software in Frontera Supercomputer.
}
\label{fig-scaling}
\end{figure}
In Fig.\ref{fig-scaling}, we compare the computational time for computing G$_0$W$_0$ band structure of the MgO system using our PyGW~\cite{PyGW_code} and Gap2~\cite{Gap2_code} code with identical input and output.  
A 8 $\times$ 8 $\times$ 8 k-point mesh with a total of 195 bands is considered for the G$_0$W$_0$ calculation. We compute G$_0$W$_0$  bands within $\pm$ 2 Ry from the Fermi energy. Both codes show linear scaling with the number of cores, however, our PyGW code is around 3 times faster than Gap2 code when using more than 80 cores, and around twice as fast for a smaller number of cores. This scaling is obtained in the Frontera supercomputer. Similar scaling is found for larger systems tested here. The reduction of the computational time is due to several improvements of the implementation: a) the efficiency of the tetrahedron method for computing the polarization in band basis is improved by precomputing common parts for all Matsubara frequencies, and more careful grouping of the diverging terms has been implemented. b) To further reduce the computational cost, we take into account that the polarization in the band basis is a real matrix, while only the matrix elements of the $\widetilde{M}$ can be complex. c) The Message Passing Interface (MPI) parallelization is here used only over bosonic momentum $\vq$ points, while OpenMP parallelization is used in internal loops over frequency, bands, and fermionic momenta.

The efficiency of the GW implementation presented here is highly advantageous for simulating metallic systems. Since the number of momentum points required for such systems is typically between one to two orders of magnitude greater than for band-insulators, the correlation self-energy becomes sensitive to the Fermi surface singularity of the single-particle Green's function. This efficiency improvement will also prove beneficial in future implementations of the diagrammatic Monte Carlo method, which systematically incorporates higher-order vertex corrections into the GW method. This necessitates a highly precise momentum mesh and accurate treatment of core states, aspects achieved only in such all-electron implementations.

\section{Conclusions}

In conclusion, we describe the implementation of GW approximation within the all-electron Linear Augmented Plane Wave framework, where we pay special attention to the metallic systems, and proper treatment of deep laying core states, as needed for the future variational diagrammatic Monte Carlo implementation. We implement both standard G$_0$W$_0$ approximation, i.e after truncating the series of self-energy to the first order in $G$ and $W$, as well as GW$_0$ algorithm, where we self-consistently compute $G$ but truncate the series in $W$ to the first order. Our improved algorithm for resolving Fermi surface singularities and frequency convolution on the Matsubara axis allows us a stable and accurate analytic continuation of imaginary axis data by Pade approximation. This is crosschecked by the contour deformation technique that avoids the need for analytic continuation.
We compute band structure and band gaps for a variety of insulators. We demonstrate the accuracy of our implementation by reproducing previous LAPW results for band insulators. We also implemented the matrix analog of G$_0$W$_0$ approximation. Here, we demonstrate \sout{ed} that the conventional diagonal approximation within the Kohn-Sham band basis is an excellent approximation, which is in contrast to the finding of Ref.~\cite{PhysRevB.104.165111}.

Surprisingly, we find that GW approximation requires an extremely dense momentum mesh for metals to converge even when tetrahedron integration is used. This is very different than in semilocal DFT approximations in which the potential is computed in real space, and therefore the convergence with momentum points is very rapid. In GW, a $16\times 16\times 16$ k-point mesh is required for reasonable convergence in the simple alkali metals such as Li, K, Na, and Mg. To test the implementation, we compute the band structures of these metallic systems and find that the converged bandwidths are slightly smaller than in LDA, for about  2-7\%. This agrees very well with the self-consistent quasi-particle GW approach.
The bandwidth in the ARPES experiment is smaller, which indicates that vertex corrections are important even in these elemental solids. The recently developed systematic approach, offered by the diagrammatic Monte Carlo method, would be very desirable to determine whether the narrowing of the bandwidth in these moderately correlated systems is purely electronic in origin, or other effects, such as interaction in the final states of ARPES experiment or the surface effects in ARPES measurements need to be considered to reproduce the experimental photoemission. Finally, we also show a substantial three-fold improvement in the speed of GW calculation compared to the previous LAPW code (gap2), on which this implementation is based.

\section{Acknowledgements} 
This research was funded by NSF DMR 2233892 and NSF OAC-2311557 and NSF OAC-2311558.
We also acknowledge support from Simons foundation, collaboration on the many electron problem. The computations were performed at the Extreme Science and Engineering Discovery Environment (XSEDE), which is supported by National Science Foundation grant number ACI-1548562, Rutgers HPC (RUPC), and the Frontera supercomputer at the Texas Advanced Computing Center (TACC) at The University of Texas at Austin, which is supported by National Science Foundation grant number OAC-1818253.

\bibliographystyle{elsarticle-num}
\bibliography{refgw}

\end{document}